\documentclass[pop,numberedappendix,twocolappendix,iop]{aeb_emulateapj_2015}
\usepackage{amsmath}
\usepackage{color}
\usepackage{longtable}
\usepackage{ulem,array}
\usepackage{grffile}

\usepackage{hyperref}
\hypersetup{
  bookmarks=false,
  pdfpagelabels=false,
  hyperfootnotes=false,
  hyperindex=false,
  pageanchor=false,
  colorlinks=true,
  linkcolor=blue,
  citecolor=blue,
  }

\usepackage{multirow}
\usepackage{amssymb}	 

\makeatletter
\newcommand{\xRightarrow}[2][]{\ext@arrow 0359\Rightarrowfill@{#1}{#2}}
\makeatother

\usepackage[mathscr]{euscript}

\newcommand{\mathbfit}[1]{\textbf{\textit{#1}}}

\renewcommand{\vec}[1]{\mathbfit{#1}}
\newcommand{\bs}[1]{\boldsymbol{#1}}
\newcommand{\A}{\mathrm{A}}
\newcommand{\K}{\dot{K}}

\newcommand{\alf}{Alfv$\acute{\text{e}}$n} %symbol for alfven with a dash :)

\DeclareSymbolFont{matha}{OML}{txmi}{m}{it}% txfonts
\DeclareMathSymbol{\varv}{\mathord}{matha}{29}

\SetSymbolFont{symbols}{bold}{OMS}{cmsy}{b}{n}
\DeclareSymbolFont{bmisymbols}{OML}{cmm}{b}{it}
\DeclareMathSymbol{\bvarv}{0}{bmisymbols}{"1D}

\usepackage{array}
\newcolumntype{M}[1]{>{\centering\arraybackslash}m{#1}}
\newcolumntype{N}{@{}m{0pt}@{}}

\begin{document}

\author{Mohamad Shalaby\altaffilmark{1} }
\author{Timon Thomas\altaffilmark{1}}
\author{Christoph Pfrommer\altaffilmark{1} }

\altaffiltext{1}{Leibniz-Institut f{\"u}r Astrophysik Potsdam (AIP), An der Sternwarte 16, 14482 Potsdam, Germany}

\email{mshalaby@live.ca}

\title{ A new cosmic ray-driven instability }
\shorttitle{A new cosmic ray-driven instability}
\shortauthors{Shalaby et al.}

\begin{abstract}
Cosmic ray (CR)-driven instabilities play a decisive role during particle acceleration at shocks and
CR propagation in galaxies and galaxy clusters. These instabilities amplify magnetic fields and
modulate CR transport so that the intrinsically collisionless CR population is tightly coupled to
the thermal plasma and provides dynamical feedback. Here, we show that CRs with a finite pitch angle
drive electromagnetic waves (along the background magnetic field) unstable on intermediate
scales between the gyro-radii of CR ions and electrons as long as CRs are drifting with a velocity less
than half of the {\alf} speed of electrons. By solving the linear dispersion relation, we show that
this new instability typically grows faster by more than an order of magnitude in comparison to the
commonly discussed resonant instability at the ion gyroscale.
We find the growth rate for this intermediate-scale instability and identify the growing modes as background ion-cyclotron modes in the frame that is comoving with the CRs.
We confirm the theoretical growth rate with a particle-in-cell (PIC) simulation and study the non-linear saturation of this instability.
We identify three important astro-physical applications of this intermediate-scale instability, which
is expected to 1.\ modulate CR transport and strengthen CR feedback in galaxies and galaxy clusters,
2.\ enable electron injection into the diffusive shock acceleration process, and 3.\ decelerate CR
escape from the sites of particle acceleration which would generate gamma-ray halos surrounding CR sources such as supernova remnants.
\end{abstract}


\section{Introduction}
\label{intro}

Propagation of CR ions through magnetized collisionless plasma occurs in many astrophysical contexts and modifies the (thermo-)dynamics of these systems in important ways.
CRs provide a source of free energy, which excites the growth of background plasma wave modes that in turn impact on their propagation in a complicated feedback loop.
In non-relativistic shocks, CRs create an effective magnetic mirror upstream of the shock front by amplifying magnetic seed fields \citep{Bell2004,Bell2005,Caprioli2014b}. This magnetic mirror scatters somewhat lower-energy ions (in comparison to the mirror-generating CRs) so that these ions can scatter back and forth across the shock front to participate in the process of diffusive shock acceleration and eventually reach very high particle energies \citep{Krymskii1977,Axford1977,Blandford1978,Bell1978a,Bell1978b,Caprioli2014a,Caprioli2014c}.

CRs in galaxies are extremely rare and reach densities that range from one part in a million to one part in a billion of the thermal background density.\footnote{Typical values of the CR-to-background density ratio in various astrophysical environments are given in Appendix~\ref{app::alphava}.} However, because CRs are a relativistic particle species, they reach energy densities that are in equipartition with magnetic and thermal pressures and thus play an important dynamical role in their evolution.
In Milky Way-like galaxies, CRs could provide a significant fraction of the perpendicular pressure that support the thermal plasma in the galactic plane \citep{Boulares1990,Boettcher+2016,Zweibel2017,Pfrommer2017b}.
The vast majority of these CRs have GeV particle energies and form a collisionless plasma that drives electromagnetic plasma waves in the background plasma (typically {\alf} waves at the ion gyroscale).
The non-linear CR-wave interaction tightly couples these collisionless CRs to the background plasma, enabling CRs to exchange energy and momentum with the background so that CRs can drive outflows from the interstellar medium \citep{2016Girichidis, 2016Simpson, 2018Farber} to larger heights in the galactic halo. This removes the fuel for star formation and thus modifies the formation and evolution of galaxies in important ways as shown in idealized simulations \citep{2012Uhlig,2013Hanasz,2014Salem,2016PakmorIII, 2017Ruszkowski,2017Wiener, 2018Jacob,2018Butsky,2019Chan,2020Dashyan}, and cosmological simulations of galaxy formation \citep{Jubelgas2008,2014SalemII,2016Salem,2020Hopkins,2020Buck}.

To incorporate such a coupling in large-scale models, effective hydrodynamic models for CR evolution and coupling to background plasmas are of prime importance \citep[see, e.g.,][]{Zweibel2017,Pfrommer2017a,Jiang2018,timon2019}. These models are derived under the assumption that CRs primarily drive resonant {\alf} waves at the ion gyroscale and scatter off of them, which leads to partial isotropization of CRs in the {\alf} wave frame and an effective streaming speed of CRs \citep{Kulsrud_Pearce-1969}. In such models, the linear growth rate of electromagnetic waves and their amplitude in the non-linear regime determine the effective scattering frequency, streaming speed, and effective diffusion coefficient of CRs. 
Moreover, this determines the CR flux that sets the forcing of the background plasma along the CR streaming direction. Thus, determining growth rate and non-linear saturation level of the excited waves is crucial in assessing the impact of CRs on the background plasma. In fact, applying this theory to MeerKAT radio observations of faint non-thermal filaments pervading the central molecular zone close to the Galactic center revealed compelling evidence that GeV CRs are mainly streaming with the local {\alf} speed \citep{Thomas2020}.

To find the linear growth rate, we have to study the linear response of the evolution equations of collisionless plasma species to a small perturbation (see Appendix~\ref{app:dispersion}).
In many astrophysical environments, the {\alf} speed is non-relativistic, i.e., $\varv_{\A}  \ll c$, where $c$ is the speed of light (see Appendix~\ref{app::alphava} for typical values of $\varv_{\A}$). 
This implies that the ion and electron gyroscales are much larger than their plasma skin depths, which in turn are much larger than their Debye length scales because $k_{\mathrm{B}} T \ll m_{i,e} c^2$, where $T$ is the background temperature, $k_\mathrm{B}$ is Boltzmann's constant, and $m_{i,e}$ is the ion or electron mass.
The background plasma temperature impacts wave modes whose wave lengths are close to the ion and electron Debye lengths. Thus, growth rates of wave modes on the ion or electron gyroscale are not affected by the background ion and electron temperature and we can model the background as a cold uniform plasma \citep[][]{Zweibel-2003}.

In order to quantify the contribution of CR ions to the dispersion relation the literature usually assumes power-law CR momentum distributions and includes CR electrons to compensate the CR current \citep{Zweibel-2003,Amato+2009,Bell2004}. This results in a complicated dispersion relation that is analytically not tractable. To simplify this dispersion relation the standard assumption is to only consider modes with oscillation frequencies much smaller than the background ion cyclotron frequency. Solving such an approximate dispersion relation\footnote{Such a configuration is unstable in the linear regime to modes that travel parallel and obliquely with respect to the background magnetic field. We focus on parallel modes since these typically have faster growth rates and thus dominate the linear evolution~\citep{Kulsrud_Pearce-1969,Zweibel2017}.} shows that CRs mainly excite {\alf} waves at the ion gyroscale \citep{Kulsrud_Pearce-1969,Zweibel-2003} and non-resonant hybrid wave modes at smaller scales (Bell modes) for high CR fluxes \citep{Bell2004,Amato+2009,Zweibel+Everett_2010}. Thus, in the context of the dispersion relation with a power-law CR momentum distribution, it is hard to assess if there are more unstable wave modes that oscillate with higher frequencies, which would be artificially suppressed in the solution of the approximate dispersion relation.

To address this point, we assume CRs with a gyrotropic momentum distribution and study the solution of the resulting dispersion relation without restricting the search to only modes with small oscillation frequency. While this gyrotropic assumption does not describe realistic CR distributions in, e.g., the interstellar medium, this allows us to solve the linear dispersion relation, and thus enables us to check the impact of the assumption of a small oscillation frequency on the solution. We will show that that the solution of the full dispersion relation reveals a new CR-driven instability, which we identify as growing background ion cyclotron modes in the comoving frame of CRs. Moreover, this ``intermediate-scale instability'' is not related to the non-resonant hybrid instability \citep{Bell2004,Bell2005} because its condition for growth requires comparably small CR fluxes, i.e., a CR drift velocity that is smaller than half of the electron {\alf} speed, which would not destabilize Bell modes.

In this context, there arises the immediate question about the astrophysical relevance of this new instability because of the artificial setting of the gyrotropic CR distribution. We note that by restricting ourselves to modes with oscillation frequencies much smaller than the background ion cyclotron frequency for a gyrotropic CR distribution, the intermediate-scale instability also disappears from the solution, just as for the power-law CR case. Hence, it is conceivable that by relaxing the  assumption of small oscillation frequencies, the intermediate-scale instability will also reveal its presence in the case of a power-law CR distribution. In fact, recently there have been two kinetic simulations published by \citet{HS+2019} (their `Lo' and `Med` simulations with a power-law CR distribution) in which the CR drift speed satisfies the condition of exciting the intermediate-scale instability and where we see a faster initial growth rate, as would be expected for this new instability.

The paper is organized as follows.
In Section~\ref{sec:disp}, we present the dispersion relation for a gyrotropic CR momentum distribution.
We present solutions of the dispersion relation in various regimes for parameters relevant to many astrophysical environments in Section~\ref{sec:sols}. 
These solutions include the new intermediate-scale instability and the previously known gyroscale, Bell, and small-scale instabilities.
We present a growth rate for the {\it new} intermediate-scale instability, and classify the regimes of dominance of all instabilities depending on the CR pitch angle and flux.
In Section~\ref{sec:sims}, we present the results from a PIC simulation using an extension of the SHARP code \citep{sharp} that demonstrates the growth of the intermediate-scale instability and study its non-linear evolution.
In Section~\ref{sec:appl}, we provide an overview of possible astrophysical applications of the intermediate-scale instability to CR propagation in galaxies and clusters (Section~\ref{sec:transport}), electron injection into the diffusive acceleration process at non-relativistic shock (Section~\ref{sec:DSA}), and CR escape from acceleration sites (Section~\ref{sec:escape}).
We conclude and  summarize our findings in Section~\ref{sec:conclusion}.
Throughout this paper, we use the ion-to-electron mass ratio $m_r = m_i/m_e=1836$ and assume the SI system of units.

\section{Linear dispersion relation with cold gyrotropic cosmic rays}
\label{sec:disp}

As we argued in the previous section, we can model the background plasma of electrons and ions as a stationary uniform cold plasma, i.e., the calculation is done in the background plasma reference frame. Namely, the background phase-space distributions are given by 
$
f_{e,i} = n_i \delta^3(\vec{u}),
$
where $\vec{u}$ is the spatial part of the four-velocity, $\delta^3$ is the Dirac delta distribution in 3D, and $n_i$ is the background number density.
Including a finite temperature in the background plasma only impacts the growth rates of instabilities at very small scales, i.e., scales of order the Debye length of ions~\citep{Reville+2008,Zweibel+Everett_2010}.
Here the background number density is assumed to be constant. Inhomogeneities in the number density can substantially change the growth rates if they are changing on scales close to the wave lengths  of the unstable modes.
However studying this theoretically, even in the linear regime, is tedious \citep[see, e.g.,][]{sim_inho_18, th_inho_20}.

Because linear analysis\footnote{See Appendix~\ref{app:dispersion} for a discussion on how the linear dispersion relation is calculated for modes parallel to the background magnetic field $B_0$.} assumes that the equilibrium configuration of all species has no net current, both CR electrons and ions have the same uniform number density $n_{\rm cr}$ and are drifting with the same mean velocity along the magnetic field, $\varv_{\rm dr}=\varv_\parallel$. That is the phase space distribution of the CR electrons is 
\begin{eqnarray}
f_{\rm cre}(\vec{x},\vec{u}) %~ d^3x ~d^3u
= \frac{ n_{\rm cr}  }{2\pi u_{\perp}}  
\delta(u_{\parallel} - \gamma_e \varv_{\rm dr}) \delta(u_{\perp})%d^3x d^3u,
~~
\end{eqnarray}
where the Lorentz factor of CR electrons is $\gamma_e = (1-\varv^2_{\rm dr}/c^2)^{-1/2}$ and $u_{\parallel} = \gamma \varv_{\parallel}$ and $u_{\perp} = \gamma \varv_{\perp}$ are defined with respect to the background magnetic field  $\vec{B}_0 = B_0 \bs{\hat{\vec{x}}}$.
The CR ions are assumed to have a gyrotropic distribution with a non-zero pitch angle so that
the phase space distribution is
\begin{eqnarray}
f_{\rm cr}(\vec{x},\vec{u}) %~ d^3x d^3u 
=  \frac{ n_{\rm cr}  }{2\pi u_{\perp}} 
\delta(u_{\parallel} - \gamma_i \varv_{\rm dr}) \delta(u_{\perp}-\gamma_i \varv_{\perp}) %d^3x d^3u,
~~
\end{eqnarray}
where $\gamma_i = \left(1- \varv^2_{\rm dr} /c^2- \varv^2_{\perp}/c^2\right)^{-1/2}$ is  the Lorentz factor of CR ions and $\varv_{\perp}$ is the perpendicular component of the CR velocity.
In our analysis here, the role of CR electrons is only to compensate the current by CR ions, i.e., we are here interested in the instabilities driven by CR ions only.
Instabilities driven by CR electrons (see, e.g.,~\citealt{Evoli+2018}), can be studied by taking gyrotropic or power-law momentum distribution for the CR electrons. However, we leave classifying the CR electron driven instabilities to future studies.

To find the linear dispersion relation that dictates the stability of the longitudinal electromagnetic wave modes, we substitute the  distribution functions given above in the dispersion relation in Equation~\eqref{eq::disp04} and obtain a relation between the complex wave frequency $\omega$ and the wave number $k=|\vec{k}|$:
\begin{eqnarray}
D^{\pm} = 0 =
1 && - \frac{k^2c^2}{\omega ^2} 
+
%%bg
\frac{\omega _i^2}{\omega  \left(-\omega  \pm \Omega _{i,0}\right)} 
+
\frac{\omega _e^2}{\omega  \left(-\omega \pm \Omega _{e,0}  \right)}
\nonumber \\ &&
+
%%CRe
\frac{\alpha  \omega _e^2 }{ \gamma _e  \omega ^2 }
\left\{
\frac{ \omega -k \varv_{\text{dr}} }
{  k \varv_{\text{dr}}-\omega \mp \Omega _{i,0} m_r/\gamma _e }
\right\}
\nonumber \\ &&
%%CRi
+
\frac{\alpha  \omega _i^2 }{ \gamma _i  \omega ^2 }
\left\{
\frac{ \omega -k \varv_{\text{dr}} }
{  k \varv_{\text{dr}}-\omega   \pm\Omega _i}
%CRiperp
-
\frac{ \varv_{\perp}^2  \left(k^2c^2-\omega ^2\right)/c^2}
{2  \left(k \varv_{\text{dr}}-\omega \pm\Omega _i \right){}^2}
\right\}.
\label{eq:dispfull}
~~~~~~~~
\end{eqnarray} 

Here, 
$\Omega_{i,0} = e B_0/m_i$ and  $\Omega_{e,0} = -e B_0/m_e = - m_r \Omega_{i,0}$ are the ion and electron cyclotron frequencies, respectively, and
$\Omega_i = \Omega_{i,0}/\gamma_i$ is the relativistic ion cyclotron frequency.
The CR-to-background number density ratio is $\alpha = n_{\rm cr}/n_i$,
$\omega^2_i= e^2 n_i/(m_i \epsilon_0) $ and $\omega^2_e = e^2 n_i/(m_e \epsilon_0)$ are the square of ion and electron plasma frequencies, respectively, $e$ is the elementary charge, $n_i$ is the background ion number density, and $\epsilon_0$ is the permittivity of free space.
The dispersion relation in Equation~\eqref{eq:dispfull} is examined by~\cite{Wu+1972} in the limit of the low density ratio, $\alpha$.
They compute the growth rate for very small scales, i.e, $k\rightarrow \infty$, and find that it is the same for any assumed frequency range. Their computed growth rates\footnote{In~\cite{Wu+1972}, the dispersion relation in Equation [10] has an error; in the numerator of the last term, $k^2$ should be replaced by $k^2-\omega^2$. This error, however, does not substantially impact the growth rates computed in the paper especially these found in the limit $k \rightarrow \infty$.
} agrees with  our findings in Section~\ref{sec:sols}.

As we show in Appendix~\ref{app:symmetry}, solving the dispersion relation  $D^{-}=0$ to find the fastest growth rates at some $k$ is obtained  by a reflection around $k=0$, and switching the sign of the real part of the solutions of $D^{+}=0$.
That is, the total growth for electromagnetic modes is twice that found by solving $D^{+}=0$, and the 2 most unstable solutions at negative and positive $k$ values have the same phase velocity $\varv_{\rm ph} = \omega_r/k$, where $\omega_r=\mathrm{Re}[\omega]$ is the real part of the wave frequency. That is, these two unstable electromagnetic wave modes at $\{k,-k\}$ are traveling in the same direction and thus are rotating in the same direction.

\section{Solutions of the linear dispersion relation  }
\label{sec:sols}

\begin{figure}
\includegraphics[width=8.4cm]{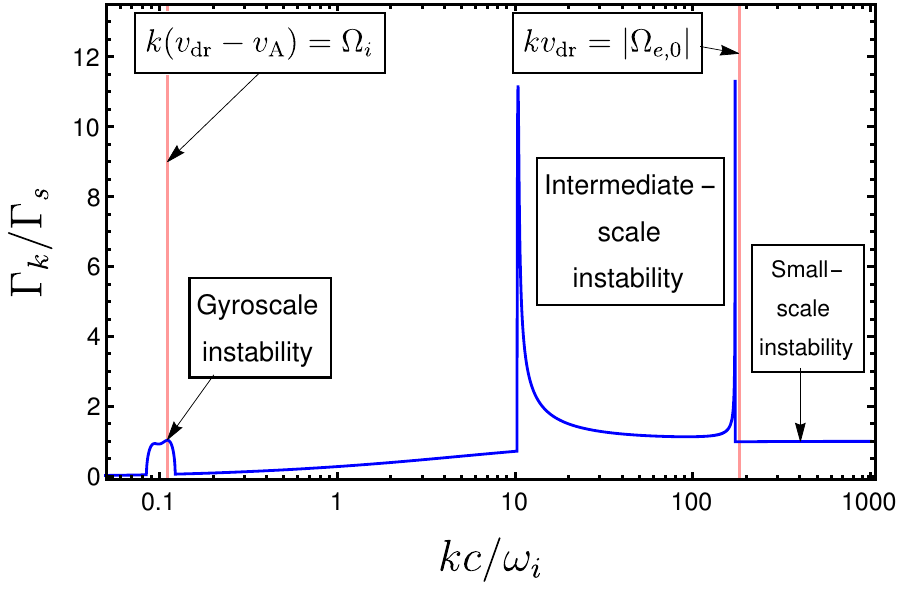}
\caption{
\label{fig:full}
Overview of the instabilities driven by gyrotropic-cold CRs with a non-zero pitch angle.
The figure shows the growth rates at various scales normalized by the small-scale instability growth rate $\Gamma_s$ of Equation~\eqref{eq::grSS}. Here, we use $ \varv_\A /c = 10^{-4} $,  $ n_{\rm cr} / n_i = 10^{-5}$, $\varv_{\rm dr} /\varv_\A = 10 $, and $\varv_{\perp} /\varv_\A = 20 $.
}
\end{figure}

In this section, we study the stability of  plasma wave modes that are parallel to the background magnetic field by searching the solutions of both dispersion relations, $D^{+}=0$ and $ D^{-}=0$ (Equation~\ref{eq:dispfull}).
Thus, the curves for the fastest growth rate are symmetric around $k=0$.

The full dispersion relation represents a polynomial of order 8, and thus it has no known closed-form solution and can only be solved numerically.
The numerical solutions $\omega(k)$ depend on four independent parameters, $\alpha$,  $\varv_{\perp}$,  $\varv_{\rm dr}$, and $\varv_\A$, where $\varv_\A = B_0/\sqrt{ \mu_0  n_i m_i} = \Omega_{i,0} c/\omega_i$ is the ion {\alf} speed, where $\mu_0$ is the permeability of free space. Thus, to facilitate studying the solutions and their dependence on various parameters, we focus on the regime of low CR density $\alpha \ll 1$ and  non-relativistic {\alf} speeds, i.e, $\varv_\A \ll c$ which is relevant for many astrophysical situations
(see Appendix~\ref{app::alphava}).

Typically the CRs have a non-zero pitch angle, i.e.,  $\varv_{\perp} \neq 0$, and for these cases, we will show in this work that they can drive a dominant intermediate-scale instability, i.e., on scales between electron and ion gyroradii.
This is the case as long as the drift velocity is less than half of the electron {\alf} speed, i.e.,
$\varv_{\rm dr} < \varv_{\A,e}/2 = \sqrt{m_r } \varv_\A/2$, where $\varv_{\A,e} = B_0/\sqrt{\mu_0 n_i m_e}$ is the electron {\alf} speed.

In Figure~\ref{fig:full}, we show the fastest growth rate 
from the dispersion relations $D^{+}=0$ and $D^{-}=0$ of Equation~\eqref{eq:dispfull} for $\varv_\A/c=10^{-4}$ and $\alpha = n_{cr}/n_i = 10^{-5}$.
In this case, CRs with finite pitch angle, i.e, $\varv_{\perp} = 2 \varv_{\rm dr}$, drive the dominant intermediate-scale instabilities which grow at a rate that is more than an order of magnitude faster than the rate by which the ion gyroscale instability grows.
Below we discuss the multi-scale instabilities driven by the gyrotopic-cold CRs shown in Figure~\ref{fig:full}.

\subsection{Small-scale and gyroscale instabilities}

\begin{figure*}
\includegraphics[width=18cm]{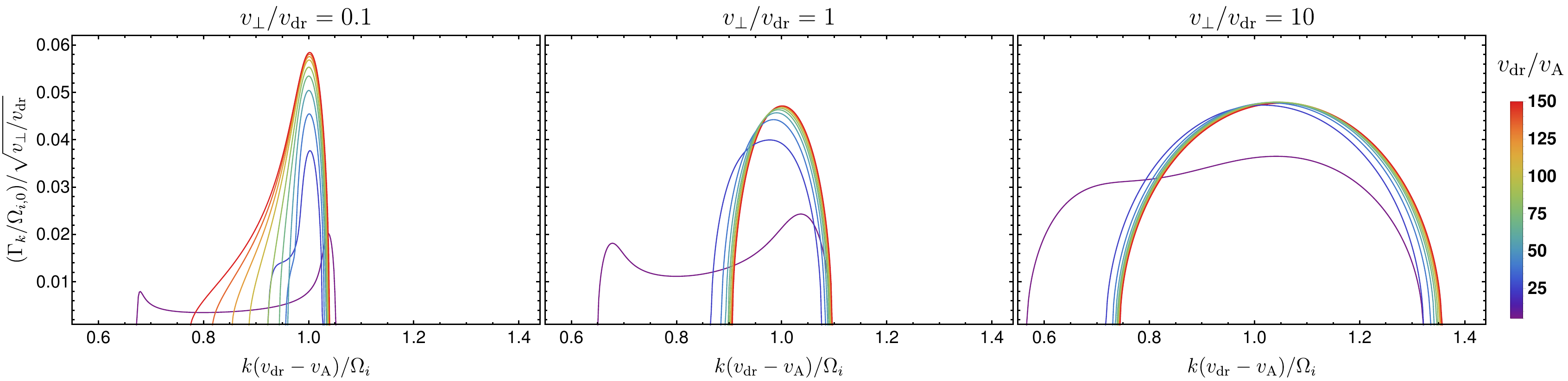}
\caption{
\label{fig:gyroinstability}
Growth rate at the gyroscale (in the {\alf} frame) for $ \varv_\A /c = 10^{-4} $ and $ n_{\rm cr} / n_i = 10^{-5}$ at various pitch-angle cosines $\mu = \varv_{\rm dr }/(\varv_{\perp}^2 + \varv_{\rm dr}^2)^{1/2}$; small (left), intermediate (middle), and large (right) pitch angle.
For each $\mu$, we show the growth rate (normalized with $ \Omega_{i,0} ~(\varv_{\perp}/\varv_{\rm dr} )^{1/2} $) for various values of the anisotropy parameter $\varv_{\rm dr}/\varv_\A$.
This shows that, for CR ions with larger pitch angles, the growth rate at the gyroscale is larger  and has larger width in $k$ space. 
}
\end{figure*}

At small scales, i.e., scales smaller than the electron gyroradius, CRs can drive these small-scale modes unstable as shown  in Figure~\ref{fig:full}.
The growth rate and the oscillation frequency of these small-scale modes are computed by~\citet{Wu+1972}, and read
\begin{eqnarray}
\label{eq::grSS}
\Gamma_s = \sqrt{ \frac{\alpha }{2} } \frac{\varv_{\perp}}{\varv_\A} \Omega_{i,0},
~~~~{\rm and}~~~~~
|\omega_{r,s} |= | k | \varv_{\rm dr} + \Omega_{i,0}.
\end{eqnarray}

That is, these modes are unstable only when $\varv_{\perp} \neq 0$.
These small-scale modes are not Bell modes~\citep{Bell2004}.
When CRs have a nonzero pitch angle, the growth rate of these small-scale instabilities is faster than that of Bell modes. However, their growth rate is maximized at scales smaller than that of the fastest unstable Bell mode.
In Section~\ref{sec:bell}, we show the distinction between small-scale and Bell unstable modes.

\begin{figure}
\includegraphics[width=8.4cm]{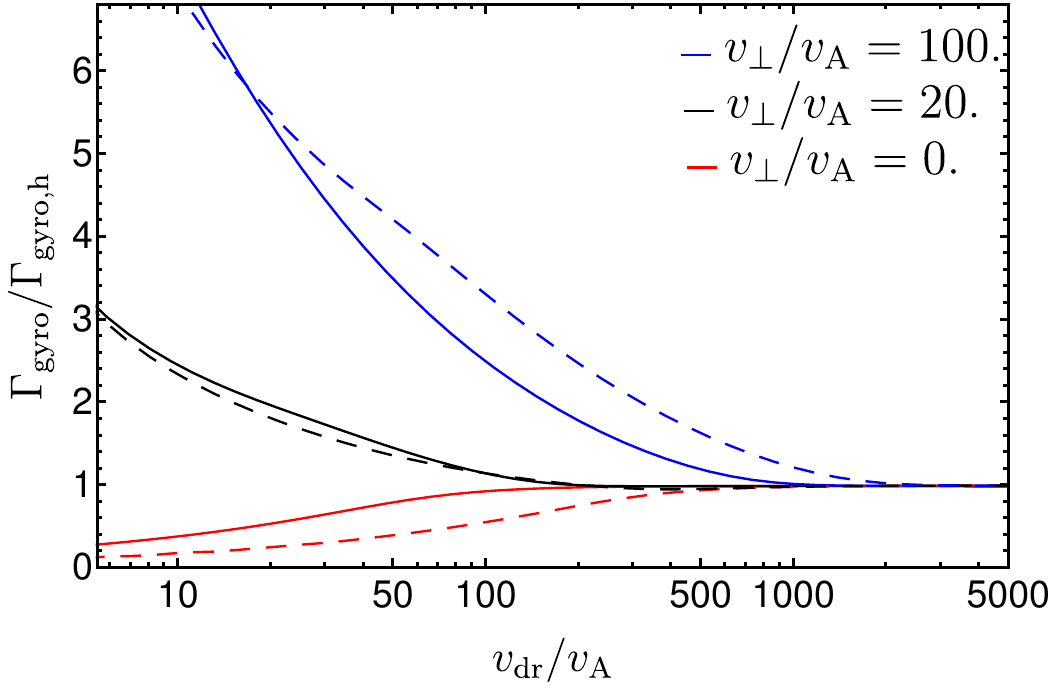}
\caption{
\label{fig:gyrodep}
The dependence of the fastest growth rate at the gyroscale, $\Gamma_{\rm gyro}$,  on  $\varv_{\rm dr}/\varv_\A$
and  $\varv_{\perp}/\varv_\A$.
Here, we use $ \varv_\A /c = 10^{-4} $,  $n_{\rm cr} / n_i = 10^{-5}$ (solid curves) and $n_{\rm cr} / n_i = 10^{-7}$ (dashed curves).
This shows a complicated dependence of $\Gamma_{\rm gyro}$ on $\varv_{\rm dr}/\varv_\A$
and  $\varv_{\perp}/\varv_\A$. However, for the high flux CRs, i.e., very large value of $\varv_{\rm dr}/\varv_\A$, the fastest growth rate asymptotically approaches  $\Gamma_{{\rm gyro},h} $ of Equation~\eqref{eq:gyroG}.
}
\end{figure}

Gyroscale instabilities are resonant instabilities, i.e., their fastest growth occurs at the gyro radius of CR ions in the {\alf} wave frame~\citep{Kulsrud_Pearce-1969}.
These instabilities exchange wave energy with CRs even when $\varv_{\perp}=0$.
However, as shown in Figure~\ref{fig:gyroinstability}, CRs with larger pitch angle have a larger rate by which they exchange energy with the unstable modes at the gyroscale, see also~\citet{Shevchenko+2002}.
The full dependence of the fastest growth rate at the gyroscale is a complicated function of 
$\varv_{\rm dr}/\varv_\A$ and  $\varv_{\perp}/\varv_\A$ as shown in Figure~\ref{fig:gyrodep}.
However, in the high flux regime, i.e., when $\varv_{\rm dr }/\varv_\A \gg 1$,
the fastest growth rate  only depends on $\alpha$, and it can be approximated by\footnote{The dependence on $\alpha$ was found by varying $\alpha$ in the range $\alpha=[10^{-7},10^{-4}]$, i.e., in the regime of $\alpha \ll 1$.}
\begin{eqnarray}
\label{eq:gyroG}
\Gamma_{{\rm gyro},h}
=
\left( \frac{ 2 	\alpha}{ 3 \gamma_i ^2 } \right)^{1/3} 
\Omega_{i,0} .
\end{eqnarray}

\subsection{Intermediate-scale instability}
\label{sec:intermediate}

\begin{figure*}
\includegraphics[width=18.6cm]{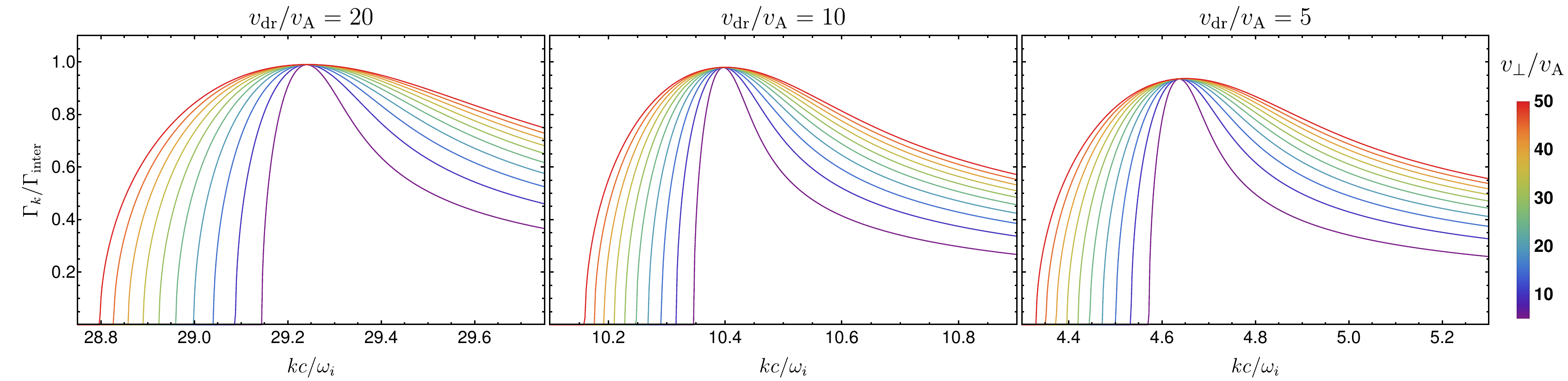}
\caption{ \label{fig:Intermediate}
Instability growth rate at intermediate scales normalized by $\Gamma_{\rm inter}$ as given in Equation~\ref{eq:IntermediateRate} which shows the proportionality
$\Gamma_{\rm inter} \propto
\left( \varv_{\rm dr} / \varv_\A  \right)^{2/3}
\left(  \varv_{\perp} / \varv_\A   \right)^{2/3}$.
The growth rate is shown for $ \varv_\A /c = 10^{-4} $ and $ n_{\rm cr} / n_i = 10^{-5}$ at various values of $\varv_{\rm dr}/\varv_\A$.
In all panels, the different colors correspond to changes in the pitch angle of CRs: the smallest pitch angles are shown in magenta and the largest pitch angles in red.  
}
\end{figure*}

As seen in Figure~\ref{fig:full}, there exists a dominant intermediate-scale instability between the scales of electron and ion gyroradii, provided the condition $\varv_{\rm dr}/\varv_\A < \sqrt{m_r}/2$ is satisfied. This instability is typically much faster than that at the gyroscale, and is typically characterized by two peaks in $k$ space as seen in Figure~\ref{fig:full}.
We compute the dependence of this instability on $\varv_{\perp}$ and $\varv_{\rm dr}$, for small values of $\alpha$ and $\varv_\A$, and find that the peak growth rate can be very well approximated by (see Figure~\ref{fig:Intermediate})
\begin{eqnarray}
\label{eq:IntermediateRate}
\frac{ \Gamma_{\rm inter} }{\Omega_{i,0} }
\approx 
\alpha^{3/4}+ 
\left( \dfrac{  \alpha  }{ 3 } \right)^{1/3}
\left( \dfrac{\varv_{\rm dr} \varv_{\perp}}{\varv_\A^2} \right)^{2/3}.
\end{eqnarray}
The growth rate of the low-$k$ peak of the intermediate-scale instability normalized by this expression is shown in Figure~\ref{fig:Intermediate} for various values of $\varv_{\perp}/\varv_\A$ and $\varv_{\rm dr}/\varv_\A$.
This also shows that CRs with larger values of $\varv_{\perp}$ have both larger growth rates and larger spectral support of the instability.\footnote{The spectral support is an important aspect when simulating the growth of the  instabilities; if the spectral support is narrow, it imposes a stringent requirement on the box size of the simulation which needs to be large enough to resolve the narrow spectral support \citep{resolution-paper}.}

For CRs with zero pitch angle, i.e., $\varv_{\perp} = 0$, this intermediate-scale instability becomes subdominant to that at the gyroscale and the spectral width becomes very narrow as found by \citet{Weidl2019}.
Already for small values of $\varv_{\perp} > 0$, the growth rate is dominated by the second term in the growth rate expression in Equation~\eqref{eq:IntermediateRate}, and it grows at a rate that is typically much faster than the growth rate of the resonant instability at the gyroscale as seen in Figure~\ref{fig:full}.
The real frequency of the intermediate-scale unstable wave modes is 
\begin{eqnarray}
|\omega_r| \sim |k| \varv_{\rm dr }  - \Omega_{i,0}.
\end{eqnarray}
Thus, these unstable modes are not whistler waves; the frequency of whistler waves~\citep{boyd} is given by $\omega \propto k^2$. 
Instead, we identify these modes as background ion-cyclotron modes in the frame that is comoving with the CRs, $|\omega_r| - |k| \varv_{\rm dr }  \sim - \Omega_{i,0}$.

\begin{figure}
\includegraphics[width=8.8cm]{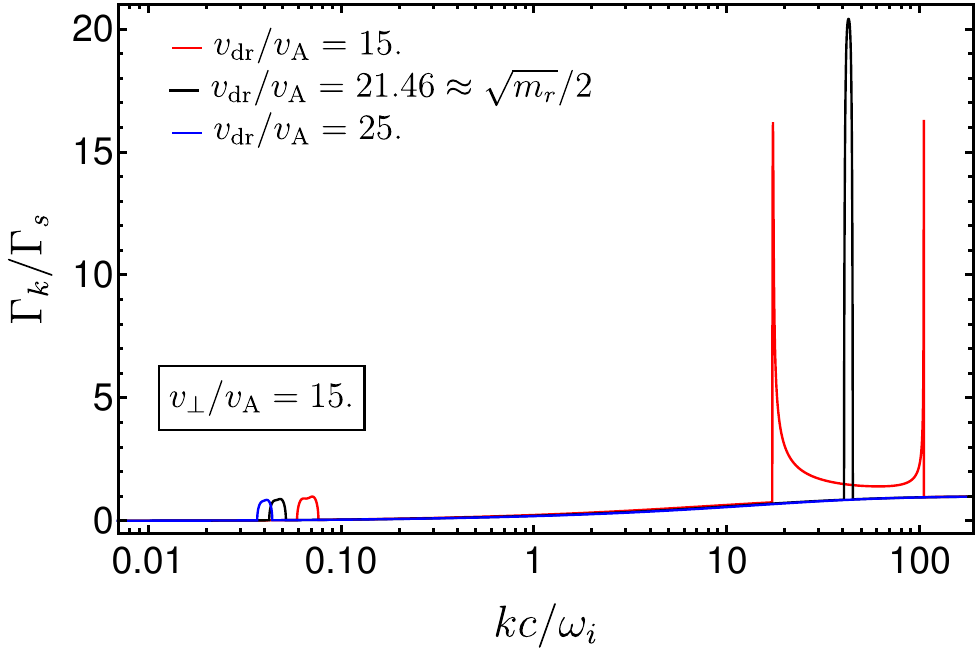}
\caption{ \label{fig:Intermediate_condition}
Impact of $\varv_{\rm dr}/\varv_\A$ on the growth rate of the  intermediate-scale instability, normalized by $\Gamma_s = \sqrt{\alpha/2} ~~ ( \varv_{\perp}/\varv_\A) ~ \Omega_{i,0}$. The instability has a double peak when $\varv_{\rm dr}/\varv_\A < \sqrt{m_r}/2$ (red). These two peaks approach each others as $\varv_{\rm dr}/\varv_\A$ increase and become a single peak when $\varv_{\rm dr}/\varv_\A \approx \sqrt{m_r}/2$ (black).
When $\varv_{\rm dr}/\varv_\A > \sqrt{m_r}/2$, the intermediate-scale instability is quenched, and only gyroscale and Bell modes are unstable (blue).
For this figure we, again, use $\alpha = 10^{-5}$, $\varv_\A/c = 10^{-4}$, and $\varv_{\perp} / \varv_\A = 15$. 
The behavior shown here is independent of the value of  $\varv_{\perp} / \varv_\A$ which only controls the spectral width of the intermediate-scale instabilities and their growth rate (see Figure~\ref{fig:Intermediate}).
}
\end{figure}

Because the two peaks of the intermediate-scale instability are placed approximately at\footnote{Exact location of the two peaks involves higher order terms of $\varv_{\rm dr}/\varv_\A$ (see Figure~\ref{fig:Intermediate}).
However, the condition for exciting the intermediate-scale instability is roughly independent of these terms, thus justifying our approximation.} 
$kc/\omega_i = \varv_{\rm dr}/\varv_\A$ and 
$k c/ \omega_i = m_r \varv_{\A}/\varv_{\rm dr} -\varv_{\rm dr}/\varv_\A $, where  $k c/ \omega_i  = m_r \varv_{\A}/\varv_{\rm dr}$ is the electron gyroscale, 
the distance in $k$ space between these two peaks is roughly  given by
\begin{eqnarray}
\frac{ \Delta k c }{ \omega_i } \approx  m_r  \frac{\varv_\A}{\varv_{\rm dr}} - 2 \frac{ \varv_{\rm dr}}{\varv_\A} 
~ \Rightarrow ~
\frac{ \Delta k \varv_{\rm dr}  }{ \Omega_{i,0} } \approx  m_r - 2 \left( \frac{\varv_{\rm dr}}{\varv_\A}  \right)^2.
~~~~~~
\end{eqnarray}
Therefore, the separation between the two-peaks decreases as $\varv_{\rm dr}/\varv_\A $ increases to the point where
the two peaks merge together when $m_r \approx 2 (\varv_{dr}/\varv_\A )^2$ (black curve in Figure~\ref{fig:Intermediate_condition}).
For  $\varv_{\rm dr}/\varv_\A > \sqrt{m_r}/2$, the intermediate-scale instability disappears completely as can be inferred from the blue curve of Figure~\ref{fig:Intermediate_condition}.

If we set $\varv_{\perp} =0$ in the dispersion relation in Equation~\eqref{eq:dispfull}, it becomes reminiscent of the approximate dispersion relation of the non-resonant hybrid instability of a power-law CR distribution \citep{Bell2004,Zweibel+Everett_2010}.
The intermediate-scale instability is artificially suppressed, i.e., absent from the solution of the dispersion relation,
if we filter out the fast dynamics,
i.e., if we assume $ \omega \ll  \Omega_i $.
For example, if we set $\varv_{\perp}=0$ in Equation~\eqref{eq:dispfull} and remove the contribution of CR electrons, i.e., drop the second line, we still recover the intermediate-scale instability in the solution to such a dispersion relation.
However, if we replace $k \varv_{\rm dr} - \omega \pm \Omega_i$ by $\pm \Omega_i$, the intermediate-scale instability is artificially suppressed in the solution.\footnote{The intermediate-scale instability is also artificially suppressed if we replace  $k \varv_{\rm dr} - \omega \pm \Omega_i$ by $k \varv_{\rm dr} \pm \Omega_i$, including the case of $\varv_{\perp} \neq 0$.}
Thus, this demonstrates that the popular assumption $\omega \ll  \Omega_i$ used in calculating the contribution of a power-law CR momentum distribution \citep{Zweibel-2003,Amato+2009,Bell2004} artificially suppresses the intermediate-scale instability.

We leave the demonstration of the existence of a dominant intermediate-scale instability in the power-law CR case for future publications.
We note, however, that two recent kinetic simulations for the power-law case presented by~\citet{HS+2019} satisfy the condition $\varv_{\rm dr}/\varv_\A < \sqrt{m_r}/2$.
Namely, the simulations `Lo' and `Med' have $\varv_{\rm dr}/\varv_\A = 1.4,~ 2.9$, respectively, and use $m_r=100$. Thus, for these two simulations, we expect a faster intermediate-scale instability during the linear evolution. Indeed, inspecting the top-panel in Figure 13 of~\citet{HS+2019} numerically demonstrates the existence of a faster growing mode in the initial evolution of this simulation (similar to our findings in Section~\ref{sec:sims}) .

A widely discussed thermal effect is ion-cyclotron thermal damping.
This is not included in our dispersion relation because background ions are assumed to have no thermal velocity ($\varv_{i,{\rm th}}=0$).
The damping  is appreciable in a small range of wave modes:
\begin{eqnarray}
  k\varv_{i,{\rm th}}\approx\epsilon \Omega_{i,0}
  \quad\Rightarrow\quad
  \frac{kc}{\omega_i} \approx \frac{\epsilon}{\sqrt{\beta}},
\end{eqnarray}
where $\epsilon=[1 ,~ 2 ]$ is the dimensionless damping range and $\beta =  \varv_{i,{\rm th}}^2 / \varv^2_\A$ is the plasma beta \citep{Zweibel+Everett_2010,Zweibel2017}.
Because the closest peak of the intermediate-scale instability to this range of wave numbers is approximately at
$
kc/\omega_i \sim \varv_{\rm dr}/\varv_\A
$, ion cyclotron thermal damping can only impact this peak if
$
\varv_{\rm dr} /\varv_\A  \approx \epsilon  / \sqrt{\beta}
$
and cannot impact the second peak.
That is, in astrophysical environments with $\beta>1$, ion-cyclotron thermal damping typically does not impact the linear growth of the intermediate-scale instability.

\subsection{Unstable Bell modes}
\label{sec:bell}

In the regime of a high CR flux, i.e., when intermediate-scale wave modes are stabilized, Bell modes~\citep{Bell2004,Bell2005} grow on scales smaller than the gyroscale.
Bell modes become unstable if
$
\alpha (\varv_{\rm dr}/\varv_\A)^2>1
$.
This was found to be the case for  a power-law CR momentum distribution \citep{Amato+2009}. Investigating the numerical solution of the dispersion relation in Equation~\eqref{eq:dispfull}, we find that this condition is also applicable to describe instability growth in the gyrotropic-cold CR case. Particularly, the growth rate of Bell modes is independent of the value of $\varv_{\perp}$.
 
Bell modes grow in a finite range in $k$ space, and their growth rate and oscillation frequency are given by~\citep{Bell2004,Zweibel+Everett_2010}
\begin{eqnarray}
\label{eq:gbell}
1 < \frac{k \varv_{\rm dr}}{\Omega_{i,0}} < \alpha \frac{\varv_{\rm dr}^2}{\varv_\A^2}
, ~
\Gamma_{\rm Bell} = \frac{\alpha}{2} \frac{\varv_{\rm dr}}{\varv_\A} \Omega_{i,0},
~
\omega_r\sim \frac{\alpha}{2} \Omega_{i,0} \ll \Gamma_{\rm Bell}.
~~~
%\nonumber \\
\end{eqnarray}

The fastest growth rate as given above is realized in approximately the middle of this instability range, that is the fastest unstable Bell mode  has a wave number~\citep{Bell2004,Zweibel+Everett_2010}
\begin{eqnarray}
\label{eq:kbell}
\frac{ k_{\rm Bell } \varv_{\rm dr} }{ \Omega_{i,0} }
=
\frac{ \alpha}{2} \frac{\varv_{\rm dr}^2}{\varv_\A^2}
~
\Rightarrow 
~
k_{\rm Bell} = \frac{ \alpha}{2} \frac{\varv_{\rm dr} }{\varv^2_\A } \Omega_{i,0}.
\end{eqnarray}

\begin{figure}
\includegraphics[width=8.4cm]{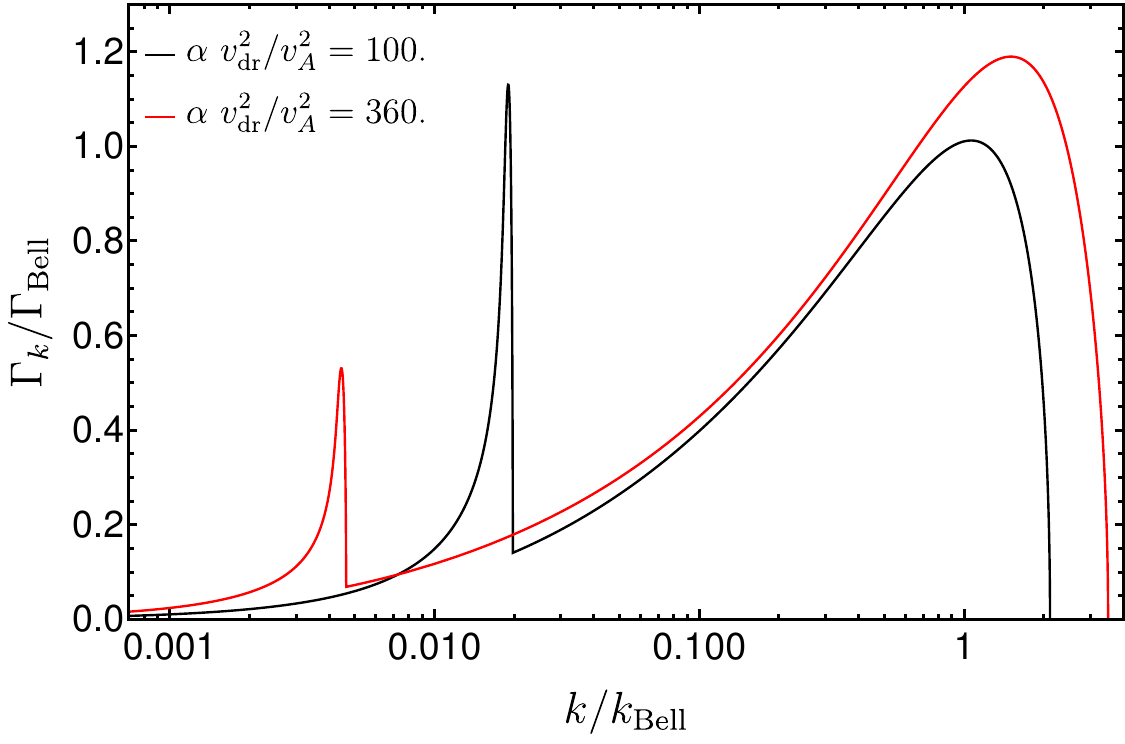}
\caption{ \label{fig:bell}
Instability growth rate (normalized by $\Gamma_{\rm Bell}$ of Equation~\ref{eq:gbell}) for Bell modes and wave modes at the gyroscale.
The growth rate of Bell modes peaks near $k=k_{\rm Bell}$ (Equation~\ref{eq:kbell}), and gyroscale modes peak at smaller values of $k$ (i.e., at larger scales).
While for both of these cases $\alpha (\varv_{\rm dr}/\varv_\A)^2>1$, Bell modes are not always growing at rates faster than the rate by which modes at the gyroscale grow.
Here, we use $ \varv_\A /c = 10^{-4} $, $ \alpha=  n_{\rm cr} / n_i = 10^{-5}$, and $\varv_{\perp}=0$.
}
\end{figure}

The condition above for the  finite range in $k$ space to be of nonzero size is that $ \alpha \varv_{\rm dr}^2/\varv_\A^2  > 1$.%, which is also the condition for the growth of the unstable Bell wave modes.
While this instability condition also applies for Bell wave modes, it is not sufficient to make their growth rate faster than that at the gyroscale.
For example, adopting the parameters $\alpha=10^{-5}$,  $\varv_{\A}=10^{-4}c$, and $\varv_{\perp}=0$ in the numerical solutions of the dispersion relation in Equation~\eqref{eq:dispfull} shows that for $ \alpha \varv_{\rm dr}^2/\varv_\A^2  <130$, the growth rate of the gyroscale instability is still faster than the growth rate of unstable Bell modes.
Figure~\ref{fig:bell} illustrates this situation for $\alpha (\varv_{\rm dr}/\varv_\A)^2=100$ (black-curve).
By contrast, for higher values of this parameter combination, i.e., for $\alpha (\varv_{\rm dr}/\varv_\A)^2=360$, the growth rate of the most unstable Bell mode is larger than the gyroscale instability growth rate (red-curve of Figure~\ref{fig:bell} ).

Indeed, because the fastest growth rate in the high-flux regime at the gyroscale is given by Equation~\eqref{eq:gyroG}, the condition for Bell modes to grow at larger rates in comparison to gyroscale modes, i.e., $\Gamma_{\rm Bell} > \Gamma_{{\rm gyro},h}$, is

\begin{eqnarray}
 \frac{\varv_{\rm dr}}{\varv_\A} > \left( \frac{16}{3  \gamma_i^2 \alpha^2  } \right)^{1/3}
{~~~~ \rm or ~~~~~~}
\alpha  \frac{\varv_{\rm dr}^2}{\varv_\A^2} > \left( \frac{256}{9 \alpha \gamma_i^4  } \right)^{1/3}.
\end{eqnarray}

For $\varv_{\perp}\neq0$, the small-scale instabilities, shown in Figure~\ref{fig:full}, can easily grow with rates that are faster than those of Bell modes.
That is, using Equations~\eqref{eq::grSS} and \eqref{eq:gbell}, $\Gamma_s > \Gamma_{\rm Bell}$ if $\varv_{\perp}> \varv_{\rm dr} \sqrt{\alpha/2}$.
However, because the fastest growth rate of the small-scale unstable modes is realized at wave lengths shorter than the electron gyroradius, this instability could be suppressed by including a finite temperature of the background electrons and ions in the dispersion relation. 
We postpone investigations of this topic to future work.

\subsection{Summary}

\begin{figure}
\includegraphics[width=8.4cm]{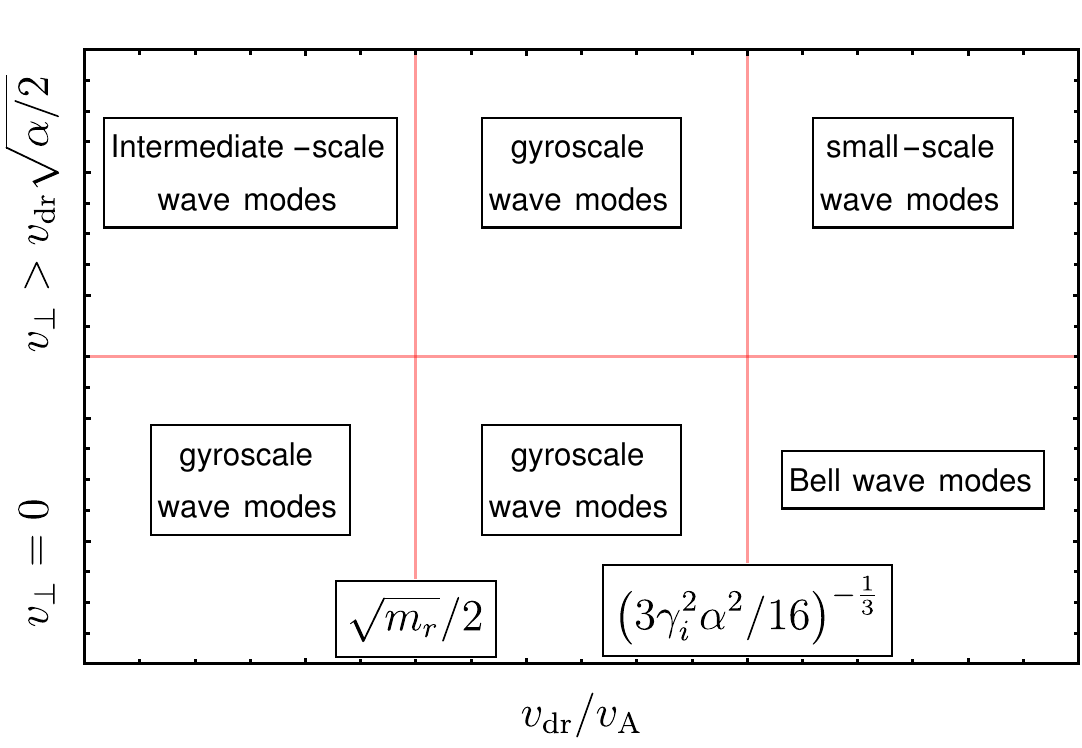}
\caption{
\label{fig:regimes}
Graphical classification of various dominant unstable modes in the linear regime driven by low density CRs with a gyrotropic cold momentum distribution. 
The fastest wave modes depend on the CR flux ($ \propto  \varv_{\rm dr}/\varv_\A$) and the pitch angle (or equivalently $\varv_{\perp}$).
Because CRs will typically have a finite $\varv_{\perp}$, the typical dominant unstable wave modes occupy the top region of the figure.
}
\end{figure}

From our discussion above, we can divide the domain of dominance of the various instabilities in terms of the CR flux, or equivalently  the value of $\varv_{\rm dr}/\varv_\A$ and whether CRs exhibit a non-zero pitch angle.
We provide a graphical summary of the dominant unstable wave modes in various regimes in Figure~\ref{fig:regimes}.
We also summarize  the growth rates, oscillation frequencies, growth scales and conditions for growth of the various instabilities driven by gyrotropic cold CR distributions in Table \ref{table:instabilities}.

\begin{deluxetable*}{ ccccc }[h!]
\tablewidth{18.1cm}
\tabletypesize{\footnotesize}
\tablecolumns{4} 
\tablecaption{
Classification of instabilities driven by gyrotropic cold CRs. 
\label{table:instabilities}
}
\tablehead{
Instability	& 
Scale of unstable wave modes  ~~~~~~~~~~&  
$\mathrm{Im} [\omega] / \Omega_{i,0} $&   
$\mathrm{Re} [\omega]$ &
Condition for growth
}
\startdata
\rule{0pt}{8pt}
Large (gyro) scales &  $k (\varv_{\rm dr} - \varv_\A) = - \Omega_ i$  & see Figure~\ref{fig:gyrodep}  &  $  k \varv_{\rm dr} + \Omega_{i,0} \sim k \varv_\A$   
&
$1 < \dfrac{\varv_{\rm dr} }{ \varv_\A } $
~
\vspace{0.2cm} 
~
\\
\rule{0pt}{12pt}
Intermediate scales\tablenotemark{a}  &
$ - m_r \Omega_{i,0}  < k \varv_{\rm dr} < \Omega_i \varv_{\rm dr}/( \varv_{\rm dr} - \varv_\A) $ & 
$ 
\alpha^{3/4}+ 
\left( \dfrac{  \alpha  }{ 3 } \right)^{1/3}
 \left( \dfrac{\varv_{\rm dr} \varv_{\perp}}{\varv_\A^2} \right)^{2/3}$
&  
$ |k| \varv_{\rm dr} - \Omega_{i,0} \sim |k| \varv_{\rm dr}$ 
 & 
$0<\dfrac{\varv_{\rm dr} }{ \varv_\A } < \sqrt{m_r} /2$  
~
\vspace{0.2cm} 
~
\\
\rule{0pt}{12pt}
Small scales
& $ |k|  \varv_{\rm dr}  > |\Omega_{e,0}| = m_r \Omega_{i,0} $  
& $\sqrt{ \dfrac{ \alpha }{2}  } \dfrac{\varv_{\perp} }{\varv_\A} $
& 
$|k| \varv_{\rm dr} + \Omega_{i,0} $
&
$ 0<\varv_{\rm dr} ~~ {\rm  and }  ~~ 0<\varv_{\perp} $
~
\vspace{0.2cm} 
~
\\
\rule{0pt}{12pt}
Bell modes\tablenotemark{b} 
& $ 1<\dfrac{ k \varv_{\rm dr}}{\Omega_i} < \alpha  \dfrac{\varv^2_{\rm dr}}{\varv^2_{\A}} $ 
&
$\dfrac{ \alpha }{2}  \dfrac{\varv_{\rm dr}}{\varv_{\A}}$
&
$\sim \dfrac{\alpha}{2} \ll \mathrm{Im}[\omega]$
&
$1 < \alpha \dfrac{\varv^2_{\rm dr}}{\varv_\A^2}  $  
\enddata
\tablenotetext{a}{For $\varv_{\perp} = 0$ and $\varv_{\rm dr} / \varv_\A  < \sqrt{m_r} /2$, a very narrow range of wave modes in the intermediate-scale regime  are unstable. However, in this case, their growth is typically slower than that on the gyroscale in agreement with results found by~\citet{Weidl2019}.}
\tablenotetext{b}{
  The growth rate of Bell modes is faster than that of the fastest unstable gyroscale wave mode if $\varv_{\rm dr}/\varv_\A > \left[ 16/(3  \gamma_i^2 \alpha^2) \right]^{1/3}$  (see Section~\ref{sec:bell}). 
}
\end{deluxetable*}

\section{Kinetic simulation of the dominant intermediate-scale instability}

\label{sec:sims}

In this Section we present a kinetic simulation using the SHARP code.
We use the SHARP-1D3V code which is an extended version of the code presented in \citet{sharp}.
In Appendix~\ref{app:1D3V}, we present the derivation of the 1D3V model (i.e., one spatial and three velocity-space dimensions) and the equations solved by the SHARP-1D3V code.

In our simulation, all background particles (ions and electrons) and CRs (ions and electrons) are modeled using macro-particles.
In PIC simulations, resolving the electron skin depth and the smaller Debye length scale of ions and electrons is important to avoid numerical heating. 
In long term simulations the usage of 5th-order interpolation, as employed by the SHARP codes, is very important to mitigate such a numerical heating~\citep{sharp}.
Numerical heating is known to have a larger impact on particles with higher energies~\citep{Lapendat2011}, and thus this aspect is very important to avoid any non-physical evolution of CR electrons and ions.
By the end of our simulation ($t = 6 \times 10^5 \omega_p^{-1}$), the total energy increases by 0.002 \% of its initial value.
That is, the energy error is 0.214 \% of the initial kinetic energy of CR ions.

\begin{figure*}
\includegraphics[width=9cm]{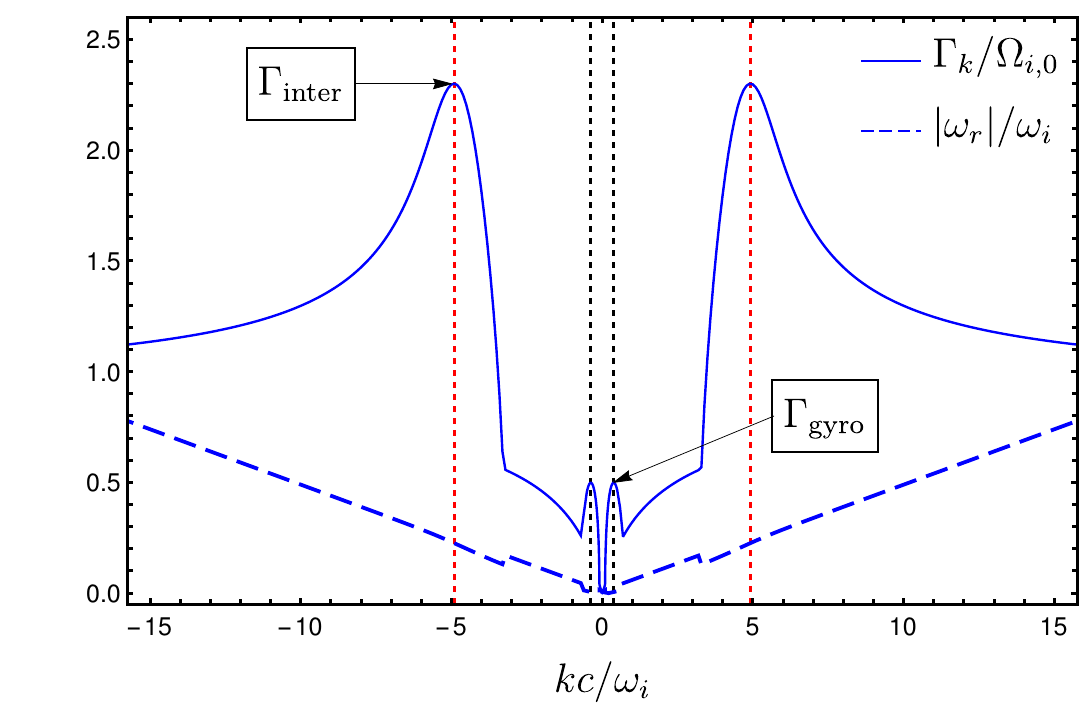}
\includegraphics[width=9cm]{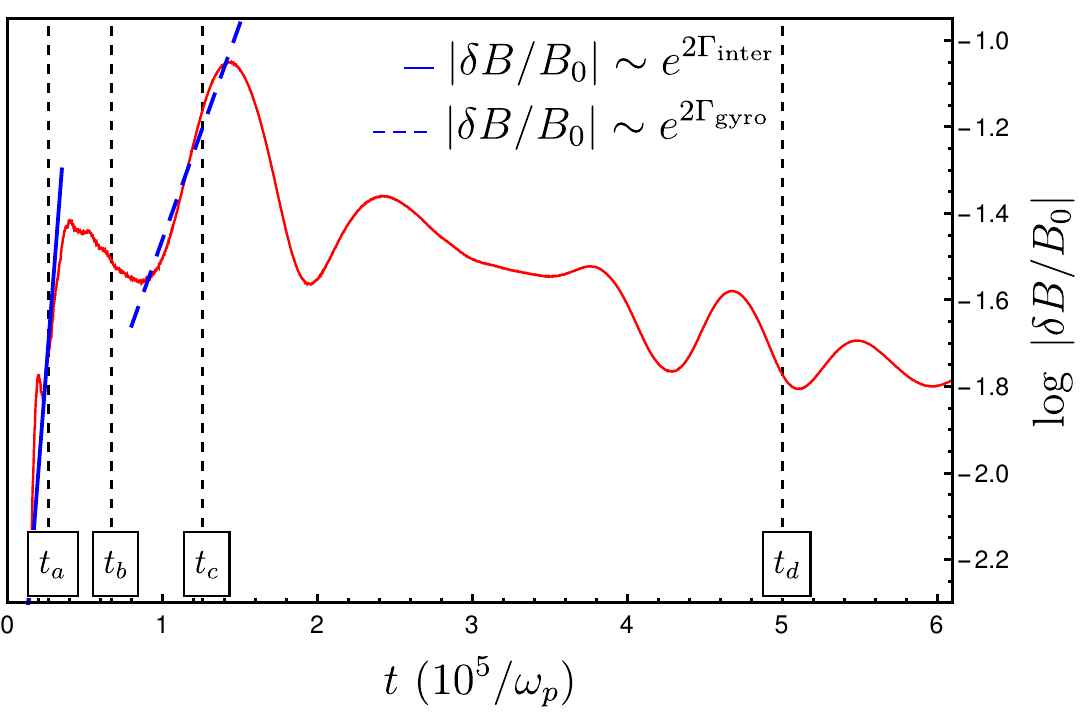}
\caption{ \label{fig:InstabilityCurve}
Left panel: the linear instability growth rate (blue) and the oscillation frequency (dashed-blue) for parameters used in our simulation:
$\alpha=10^{-2}$, $\varv_\A / c= 10^{-2}$, $\varv_{\rm dr} = 5 \varv_\A$, and $\varv_{\perp} = 13.1 \varv_\A$.
Red-dashed lines indicate the fastest intermediate-scale unstable wave mode and black-dashed lines indicate the gyro scale.
Right panel: the growth history of transverse magnetic field fluctuations in our simulation. Here, we label four times that mark important stages during the evolution: $t_a$ , $t_b$ , $t_c$ , and $t_d$.
Blue (dashed-blue) line indicates the growth rate of intermediate-scale (gyroscale) instability. 
}
\end{figure*}

In order to reduce the computational cost of the simulation we use a higher value of the density contrast $\alpha=10^{-2}$, which yields a larger growth rate. We also use a higher value of $\varv_\A=10^{-2}c$ to reduce the separation between the ion gyroradius and the electron skin depth. For CR ions, we  use $\varv_{\perp} = 13.1 \varv_\A$ and $\varv_{\rm dr} = 5 \varv_\A$.
By solving the dispersion relation numerically for these parameters, we find the fastest growth rate  (shown in the left panel of Figure~\ref{fig:InstabilityCurve}).
Our choice of $\varv_\A$ and $\alpha$ decreases the ratio of the intermediate-scale to gyroscale growth rates when compared to, e.g., Figure~\ref{fig:full}.
However, the intermediate-scale instability is still faster than that at the gyroscale.
In the left panel of Figure~\ref{fig:InstabilityCurve}, the red-dashed (black-dashed) lines indicate the scale of the fastest unstable wave modes at intermediate scales (the gyroscale).

From the numerical solutions of the dispersion relation, we also compute the phase velocity of the most unstable intermediate-scale  and gyroscale  wave modes, and find 
\begin{eqnarray}
\label{eq:vph}
\varv_{\rm ph,\,inter}
&=&
0.046 ~~~c \approx \varv_{\rm dr},~~~{\rm and }\\
\varv_{\rm ph,\,gyro}
&=&
0.0134 ~ c \approx \varv_\A.
\end{eqnarray}

In our simulation we resolve the plasma skin depth ($c/\omega_p$) by 10 cells, and our time step is $0.04 ~ \omega_p^{-1}$, where $\omega^2_p = (\alpha+1) ( \omega^2_i + \omega^2_e) = (\alpha+1)(m_r+1) (\Omega_{i,0}\,c/\varv_\A)^2 $ is the square of the total plasma frequency of all species, and we use a realistic ion-to-electron mass ratio\footnote{All gyrotropic kinetic simulations presented by \citet{HS+2019} use a reduced mass ratio, $m_r=100$. Computing the linear growth rates for all of those simulations (similar to the growth rate shown in the left panel of Figure~\ref{fig:InstabilityCurve}), we see that the intermediate-scale instability is artificially suppressed owing to the reduced mass ratio, which would not be the case for a realistic mass ratio.}, $m_r=1836$.

The background electrons and ions are uniformly initialized on the grid and their momenta follow a Gaussian distribution with the same physical temperature
$T = 10^{-4} m_i c^2/k_\mathrm{B}$.
CR electrons and ions are uniformly initialized on the grid with parallel velocity $\varv_{\rm dr} = 5 \varv_\A$ in the direction of the background magnetic field along the positive $x$ direction.
For CR electrons the perpendicular velocities are initially set to zero, i.e., $\varv_y=\varv_z=0$. 
By contrast, the perpendicular velocities of CR ions are randomly chosen such that $ (\varv_y^2 + \varv_z^2)^{1/2} = \varv_{\perp} = 13.1 \varv_\A$.
Background ions and electrons are represented by $2,500$ macro particles per computational cell per species, while CR ions and electrons are represented by $25$ macro particles per species.
The initial momentum distribution for CR ions is indicated with a star in the various panels of Figure~\ref{fig:phaseSpace}.

The computational domain is comprised of $109,715$ cells, which yields the box size 
$L = 10971.5 ~ c/\omega_p \sim 10.14 ~ r_g$, where $r_g = 2 \pi (\varv_{\rm dr}-\varv_\A) /\Omega_i $ is the wave length of the most unstable wave mode at the gyroscale.

\subsection{Growth of transverse magnetic fluctuations}

\begin{figure}
\includegraphics[width=8.4cm]{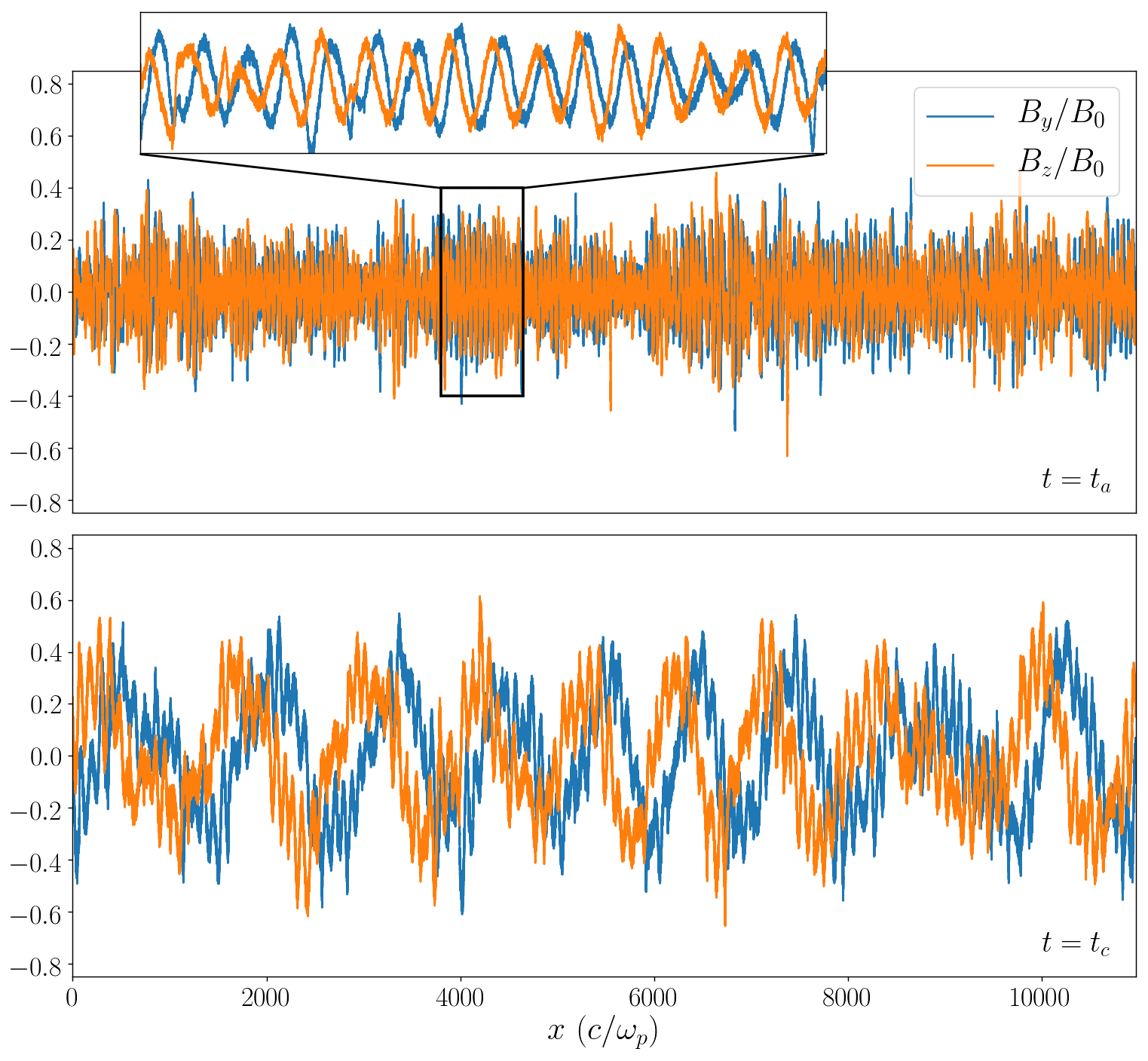}
\caption{
\label{fig:Bx}
Transverse magnetic field fluctuations during linear growth of the intermediate-scale instability (top) and during linear growth of the gyroscale instability (bottom). The times $t_a$ and $t_c$ are marked by vertical black-dashed lines in the instability growth history (right panel of Figure~\ref{fig:InstabilityCurve}).
}
\end{figure}

\begin{figure*}
\includegraphics[width=18cm]{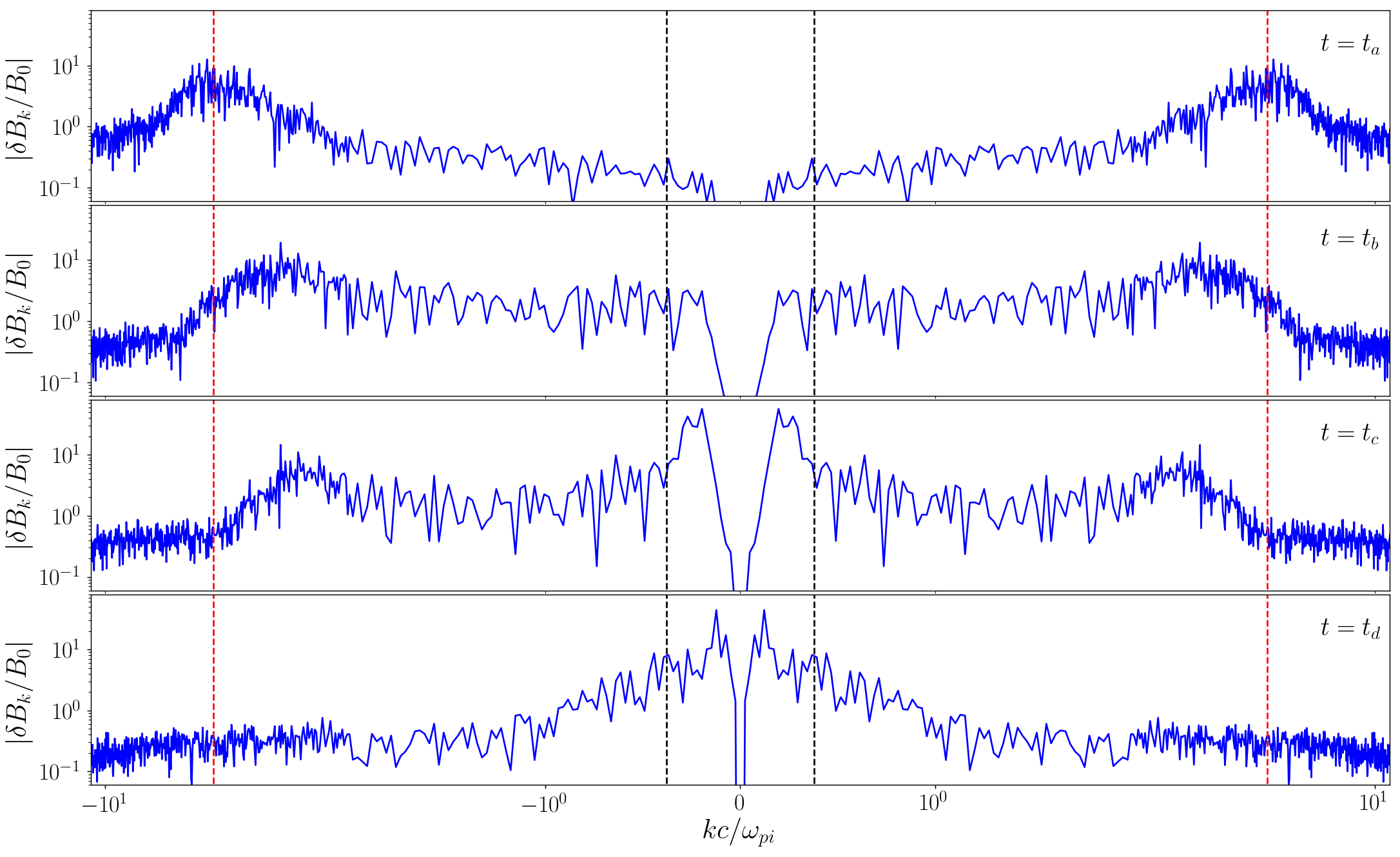}
\caption{
\label{fig:Bk}
Fourier transform of excited transverse magnetic field.
Red-dashed (black-dashed) line indicates the position of the fastest wave modes from the intermediate-scale (gyroscale) instability (see left panel of Figure~\ref{fig:InstabilityCurve}),
and we use $|\delta B_k| = \delta B_{y,k} +\delta B_{z,k}$.
The four panels show different stages during the evolution of the system; the time $t_a$, $t_b$, $t_c$, and $t_d$ are marked by black-dashed lines in the right panel of Figure~\ref{fig:InstabilityCurve}.
}
\end{figure*}

The right panel of Figure~\ref{fig:InstabilityCurve} shows the evolution of the transverse magnetic field fluctuations indicating two distinct stages during the linear evolution.
During the first (second) stage, the magnetic field fluctuations grow at a rate that agrees very well with the fastest growth rate of the intermediate-scale (gyroscale) instability.
We label four stages of importance during the evolution;
the first time ($t=t_a$) corresponds to the linear evolution of the intermediate-scale instability,
the second ($t=t_b$) corresponds to  non-linear saturation of the intermediate-scale instability, 
the third time ($t=t_c$) corresponds to the linear evolution of the gyroscale instability, 
and the fourth ($t=t_d$) corresponds to  non-linear saturation of the gyroscale instability.

In Figure~\ref{fig:Bx}, we show the spatial configuration of the transverse magnetic field components. The top panel shows the fields during the growth of the intermediate-scale instability (at $t=t_a$), while in the
bottom panel, we show the  fields during the growth of the gyroscale instability (at $t=t_c$).
This demonstrates that electromagnetic waves grow in both cases but on different scales.

We show the wave numbers of amplified magnetic field fluctuations in Figure~\ref{fig:Bk} at various times ($t_a$, $t_b$, $t_c$, and $t_c$), where
$|\delta B_k| = \delta B_{y,k} +\delta B_{z,k}$.
At $t=t_a$, the evolution is dominated by the growth of the dominant intermediate-scale instability, and we see a clear peak in the power of excited wave modes  around the red-dashed line which indicates the scale of the fastest intermediate-scale unstable wave mode as show in Figure~\ref{fig:InstabilityCurve}.
The saturation of the intermediate-scale instability occurs by cascading power to longer wave length
as shown in the second panel of Figure~\ref{fig:Bk} ($t=t_b$).

After non-linear saturation of the intermediate-scale instability, another stage of linear growth starts. We clearly see growth of unstable modes at a wave length comparable to the gyroscale of the initial configuration (indicated by black-dashed lines) in the third panel of Figure~\ref{fig:Bk} ($t=t_c$).
The peak growth occurs at a slightly larger scale in comparison to the wave length of the gyroscale $r_g =  2 \pi (\varv_{\rm dr}-\varv_\A) /\Omega_i$. The reason for this deviation is the evolution of the CR velocity-space distribution from its initial configuration to just before the second growth stage as can be seen in the top-left panel of Figure~\ref{fig:phaseSpace}.
Since the size of the computation domain is $L\sim 10 ~r_g$, this is also manifested in the bottom panel of Figure~\ref{fig:Bx}, where we see nine wave lengths in the computation domain instead of the expected ten waves.  
After non-linear saturation of the gyroscale instability, the cascade to longer wave lengths continues (bottom panel of Figure~\ref{fig:Bk}).
However, whether such trend of cascading power to longer wave lengths continues cannot be addressed within the simulation because at the end of the simulation, the largest power accumulates on a scale that corresponds to about $0.33$ of the computational domain length which could hinder further cascading if we run our simulation further.
Moreover, we caution that mode-mode coupling and thus the cascading behaviour strongly depends on the dimensionality of the problem: if we were to run simulations in two or three spatial dimensions we could obtain a different non-linear evolution.

\subsection{Scattering of CRs in velocity space}

\begin{figure*}
\includegraphics[width=18cm]{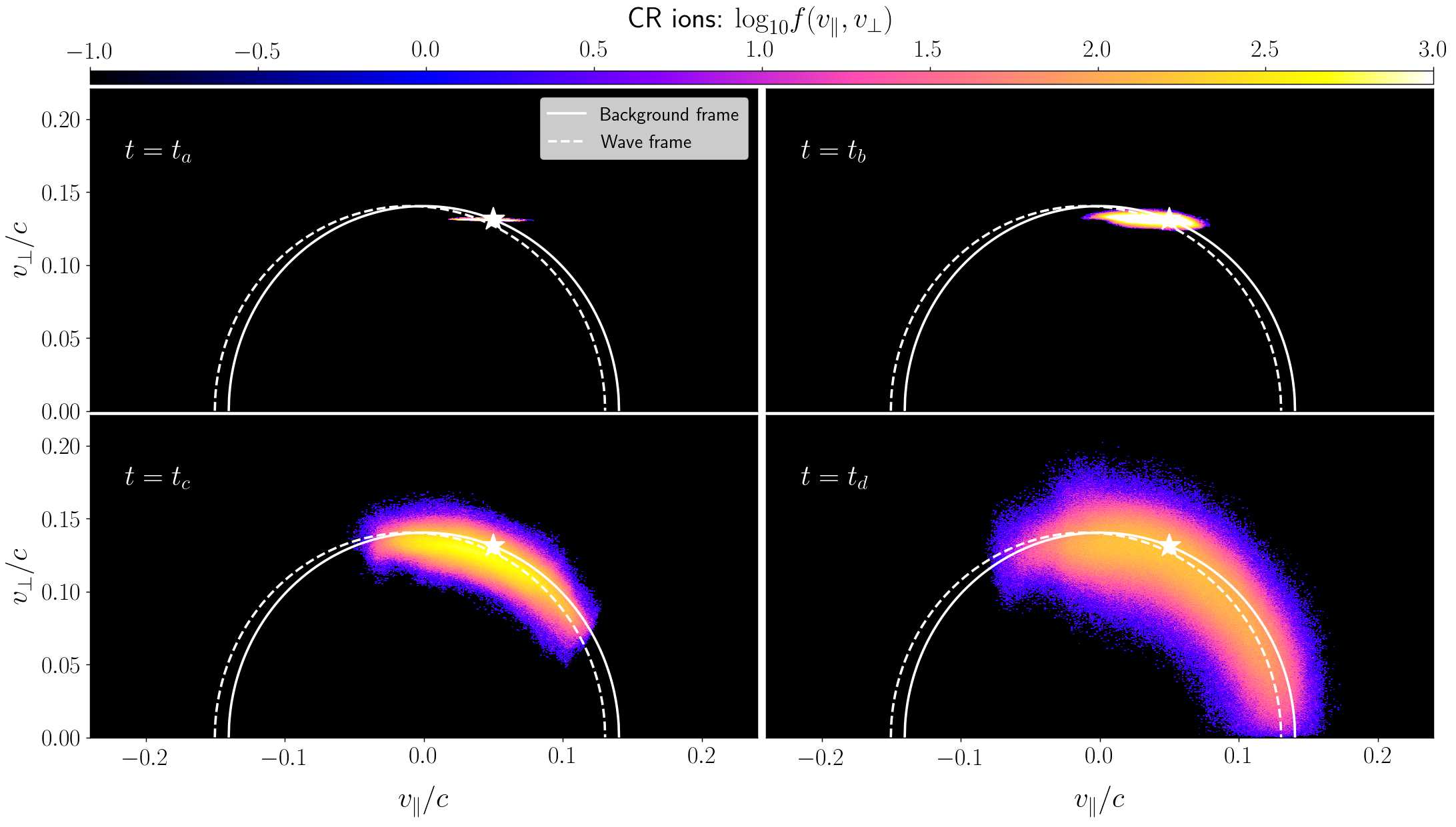}
\caption{
\label{fig:phaseSpace}
Velocity-space distributions of CR ions at different times marked by black-dashed lines in the right panel of Figure~\ref{fig:InstabilityCurve}.
White (white-dashed) semi circles indicate surfaces of constant energy in the background ({\alf} wave) frame and the star denotes the initial gyrotropic CR ion distribution.
}
\end{figure*}

In the linear regime, the rate of change of the parallel and perpendicular energy of a particle with charge $q_s$ due to its interaction with an electromagnetic wave mode with phase speed, $\varv_{\rm ph}$, is given by (see Appendix~\ref{app:scattering})
\begin{eqnarray}
\K_{\parallel}
&=&
\frac{m_s}{2} \frac{d \varv^2_x }{dt}
=
q_s \varv_x (\varv_y  B_z -  \varv_z B_y ),
\nonumber \\
\K_{\perp}
&=&
\frac{m_s}{2} \frac{d \varv^2_\perp }{dt}
=
- q_s \left[ (\varv_x - \varv_{\rm ph} )\varv_y  B_z   -( \varv_x - \varv_{\rm ph} )  \varv_z B_y \right].
\label{eq:energyrate}
\end{eqnarray}
For the intermediate-scale instability we have $\varv_x \approx \varv_{\rm dr} \approx \varv_{\rm ph}$, which implies $\K_{\perp} \approx 0 $ and $\K_{\parallel} \neq 0$. Therefore, during the linear growth phase of the intermediate-scale instability we expect a spread in parallel CR velocity while maintaining $\varv_{\perp}=\mathrm{const}$.

For the gyroscale instability, particles in the {\alf}-wave frame have $\varv_{\rm ph}=0$. Hence, we obtain $\K_{\parallel} = - \K_{\perp}$, i.e., the gyroscale instability is expected to scatter CRs along lines of constant energy in the {\alf}-wave frame. 
This is a manifestation of the fact that in the {\alf} frame, magnetic fields are stationary so that the electric field vanishes and particles can only scatter in pitch angle while conserving their energy.

\begin{figure}
\includegraphics[width=8.7cm]{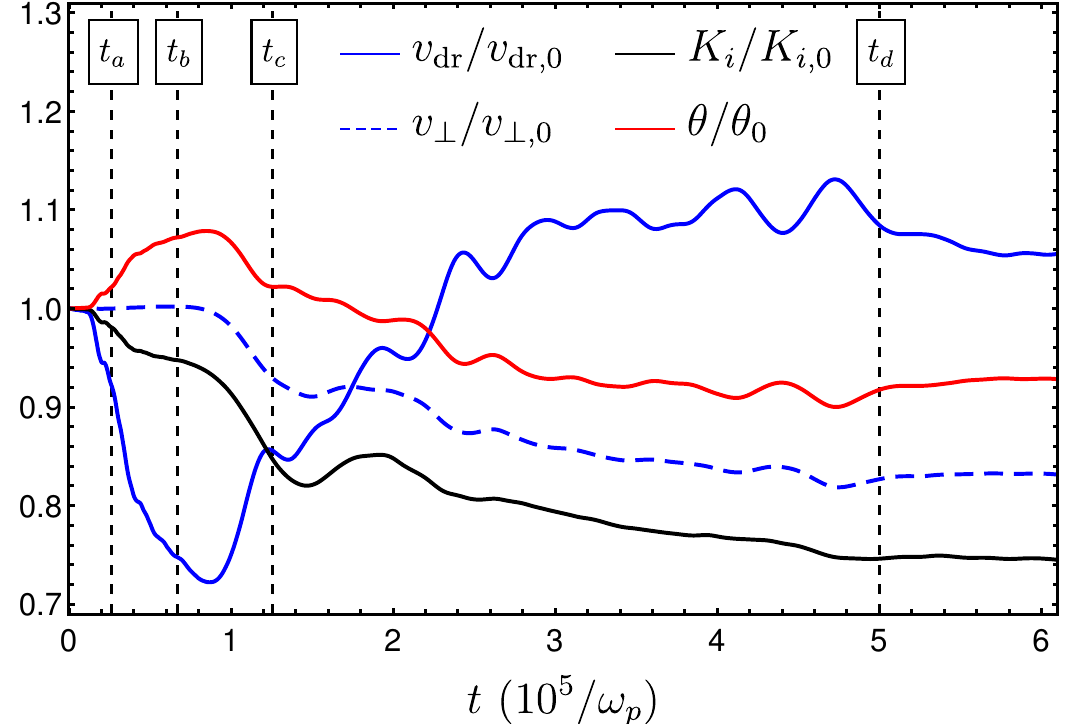}
\caption{
\label{fig:vdrvp}
Evolution of CR ion velocities.
The average parallel ($\varv_{\rm dr}$) and perpendicular ($\varv_{\perp}$) mean velocity evolution is shown with blue and dashed-blue curves, respectively.
Here, $\varv_{\rm dr,0}$ and $\varv_{\perp,0}$ are the initial parallel and perpendicular CRs velocities, respectively.
The black curve shows the evolution of the kinetic energy of CR ions, $K_i = 0.5~m_i \left( \varv_{\perp}^2 + \varv_{\rm dr}^2 \right)$, and $K_{i,0}$ is the initial CR ion kinetic energy.
The red curve shows the evolution of the pitch angle, $\theta = \cos^{-1} \left[ \varv_{\rm dr}/(\varv_{\perp}^2 + \varv_{\rm dr}^2)^{1/2} \right]$, normalized to its initial value $\theta_0 = 0.384 ~\pi$.
}
\end{figure}

The impact of these instabilities on CR ion velocities and their total kinetic energy during the evolution of our simulation is shown in Figures~\ref{fig:phaseSpace} and \ref{fig:vdrvp}.
During linear growth of the intermediate-scale instability, the average perpendicular velocity remains approximately constant while we observe an increasing parallel velocity dispersion, just as predicted, see the top-left panel of Figure~\ref{fig:phaseSpace}. This corresponds to a decrease in the average drift velocity.
During the non-linear stages of this instability, the power starts to cascade to longer wave length
and we start to see a small spread in the perpendicular velocity while the spread in drift velocities and the decrease of its average value continues.

During the linear growth stage of the gyroscale instability, i.e., near $t=t_c$, most of the power is at wave lengths comparable to the gyroscale. Because the phase velocity of these wave modes is approximately $\varv_\A$ in the lab frame (i.e., $\varv_{\rm ph}=0$ in the {\alf}-wave frame) we see scattering along the line of constant energy in the wave-frame (dashed-white lines in Figure~\ref{fig:phaseSpace}), confirming our discussion following  Equation~\eqref{eq:energyrate}.
This is also manifested in the approximately constant CR ion energy between $t=t_c$ and 
$t=2\times10^2 \omega_p^{-1}$ (cf.\ the black-curve in Figure~\ref{fig:vdrvp}). 
During the growth of gyroscale wave modes, we see a significant spread in both parallel and perpendicular velocities of CR ions (bottom-left panel of Figure~\ref{fig:phaseSpace}).

After non-linear saturation of the gyroscale instability, the cascade of power to longer wave lengths continues, which increases the spread in CR ion velocities as can be seen in the bottom-right panel of Figure~\ref{fig:phaseSpace}. This corresponds to a continued decrease of the average kinetic energy of CR ions as we can see from Figure~\ref{fig:vdrvp}. Simultaneously, the power in the magnetic field is approximately constant or even slightly decreasing.
This is a clear indication that at the non-linear stage, CR ions continue to lose energy to the excited wave modes, and another (damping) mechanism is responsible for moving this energy from the waves to the background plasma.
By the end of the simulation the total energy of background ions and electrons have increased by $\sim 10$\,\% and $\sim 3.4$\,\% of their initial thermal values, respectively.
During these later stages, the average CR  drift velocity increases by about 8\,\% of its initial value since the majority of the spread in drift velocity is along the initial drift direction.
An animation for the full evolution during the simulation can be found \href{https://www.youtube.com/watch?v=yP2dBCPOQYc&t=4s&ab_channel=MohamadShalaby}{here}.

\subsection{Implications and limitations of our kinetic simulation}

The numerical simulation successfully shows that the intermediate-scale wave modes that we identified by solving the linear dispersion relation, grow at the expected rate and also at the expected wave length.
The values of the parameters $\alpha$ and $\varv_\A$ imply a ratio of $\Gamma_{\rm inter}/\Gamma_{\rm gyro} = 4.6$. This is small in comparison to the value $\Gamma_{\rm inter}/\Gamma_{\rm gyro} \sim 20$, which we obtain with more realistic parameters (see, e.g., Figure~\ref{fig:full}).
That is, in this simulation the dominance of the intermediate-scale instability is reduced, and despite this, we observe significant magnetic field amplification as a result of the intermediate-scale unstable wave modes.

While our 1D3V setup is sufficient to demonstrate the linear evolution of the system, the non-linear evolution presented above could significantly change in two or three spatial dimensions. It is well known that the turbulent cascade is substantially modified in higher spatial dimensions~\citep[see, e.g.,][]{Muller-2000}. Because non-linear saturation of the instabilities in our simulation occurs via cascading energy to longer wave lengths, the realistic non-linear evolution of this problem may not be fully captured in our  simulation.

Moreover, it is usually assumed that magnetic energy generated at the gyroscale remains at the same scale while instability growth is countered by various damping mechanisms \citep{Zweibel2017,Jiang2018,timon2019}.
However, during non-linear saturation of the gyroscale instability in our simulation, the magnetic field does not remain at the fastest unstable scale, questioning the usual assumption.
However, as we point out above, because mode-mode coupling and thus wave cascading works differently in higher spatial dimensions, our simulation may not disprove the usual assumption.
We postpone a study  of the non-linear evolution of the intermediate-scale instability in higher spatial dimensions to future work.

\section{Applications} \label{sec:appl}

\subsection{CR transport in galaxies and clusters}
\label{sec:transport}

In the self-confinement picture, CRs are assumed to be only scattered by wave modes driven by propagating CRs, i.e., any external magnetic turbulence is ignored.
In this picture, the evolution of the CR energy density, $\varepsilon_c$, and pressure, $P_c$, are governed by the so called classical theory of CR hydrodynamics (CCRH)~\citep[see, e.g.,][for a recent review]{Zweibel2017}.
While, the derivation of this theory typically assumes that CRs are scattered by {\alf} waves, formally a similar derivation is needed to assess the impact of the intermediate-scale instability.
However, we can get a qualitative estimate for its effect by noting that in CCRH, the evolution of the CR energy density is given by
\begin{eqnarray}
&&
\frac{d \varepsilon_c}{dt} + \nabla \cdot \left[ \vec{W} (\varepsilon_c + P_c) 
- 
\boldsymbol{   \kappa }
\cdot \nabla \varepsilon_c\right]
=
\vec{W} \cdot \nabla P_c,
\\
&&
\label{eq:Gr}
\vec{W} \cdot \nabla P_c
=
-
2 \int d\omega dk \Gamma(\omega, k) I(\omega,k),
\\
&&
| \boldsymbol{   \kappa } | \sim {\kappa}_{\parallel} \sim \frac{c^2}{2} 
\left\langle 
\frac{1 - \mu^2}{\nu_{+} + \nu_{-}}
\right\rangle,
\end{eqnarray}
where $\vec{W}$ is the effective streaming speed of CRs, 
$\Gamma(\omega, k)$ is the growth rate of the wave mode $k$ that oscillates with frequency $\sim \omega$, and $I(\omega,k)$ is the amplitude of the wave mode. Typically it is assumed that the wave amplitudes are in steady state, i.e., constant, because wave growth is balanced by wave damping processes.
The diffusion tensor, $ \boldsymbol{   \kappa }$, is typically dominated by the parallel diffusion coefficient $\kappa_{\parallel}$, which depends on the effective scattering frequency  of the rightward and leftward propagating wave modes, $\nu_{\pm}$~\citep{Zweibel2017}.

From Equation~\eqref{eq:Gr}, we can already see that the pressure gradient along the streaming velocity directly depends on the growth rate of the unstable wave modes.
Typically, $\Gamma_{\rm inter}/\Gamma_{\rm gyro} \sim 20$, which implies that the CR pressure gradient created by scattering CRs off of the intermediate-scale unstable wave modes would be more than an order of magnitude larger than what is typically assumed in the CCRH picture.
Moreover, the substantially larger instability growth rate could also manifest itself in a different scattering frequency and thus a different effective streaming speed and parallel diffusion coefficient.
That is, additionally accounting for the intermediate-scale instability  could substantially change the evolution of the CR energy density and its dynamical impact on the surrounding plasma.

Such a larger induced CR pressure gradient could greatly enhance the strengths of galactic winds and thus mitigate the star formation rates in galaxies by moving gas to large radii as shown in simulations of galaxy formation (see Section~\ref{intro}).

This could also have a substantial impact on the physical evolution of the intracluster medium in galaxy clusters. Approximately half of all clusters have a cooling time of the central X-ray emitting gas of less than 1~Gyr. In the absence of any heating process, these hot gaseous atmospheres are expected to cool and to form stars at rates up to several hundred $M_\odot~\mathrm{yr}^{-1}$ \citep[see][]{Peterson2006}. Instead, energy delivered by radio lobes that are inflated by jets of active galactic nuclei appear to balance the cooling through a self-regulated feedback loop \citep{McNamara2007}.
A promising idea for the heating process are CRs that are accelerated in internal shocks of these jets. After escaping from the lobes and streaming along the magnetic field, they can excite electromagnetic waves via the gyro-resonant and intermediate-scale instabilities. Damping these waves heats the surrounding cooling gas \citep{Loewenstein1991,guo2008,Fujita2012,pfrommer2013,jacob2017a,jacob2017b,ruszkowski2017,ehlert2018} at a larger rate in comparison to the classical picture, which only considers the gyro-resonant instability  because of the increased CR pressure gradient (Equation~\ref{eq:Gr}).

\subsection{Injection of thermal electrons into the diffusive shock acceleration process}
\label{sec:DSA}

In non-relativistic electron-ion shocks, ion-gyroscale instabilities grow at the shock transition to mediate the shock, yielding a shock width that is typically of order the ion gyroradius.
The ions always have a very anisotropic distribution in the frame of the plasma after crossing the shock transition so that they can become energized upon isotropizing in the corresponding (up- or downstream) frames by virtue of momentum conservation \citep{Bell1978a,Caprioli2015}.
Thermal electrons have a smaller gyroradius by $ m_r $. The electron injection problem arises because thermal electrons cannot cross the shock transition region in one gyro orbit, but instead randomly walk in the strong electrostatic shock potential along the local mean magnetic field~\citep{Bykov+1999}.
This causes them to scatter and to partially isotropize in the transition region before exiting it on the other side. In contrast to ion injection, electrons cannot experience a similar injection as they exit the transition region with a partially isotropized distribution which would not allow them to experience acceleration via diffusive shock acceleration~\citep{Malkov+2001}.

Instabilities shown in Figure~\ref{fig:full} could solve this problem.
Since the drifting ions typically have finite pitch angle, small-scale unstable wave modes on scales close to and smaller than the electron gyroscale grow with a similar rate to that at the gyroscale.
These can scatter and accelerate thermal electrons.
Once the accelerated electrons have a gyroradius in the intermediate-scale instability range, they are scattered very strongly by the larger amplitude wave modes driven by the dominant intermediate-scale instability.
That is, this instability drives wave modes that continue to accelerate electrons at quasi-parallel shocks until their gyroradius becomes comparable to that of ions at which point they could be scattered across the shock front.
The condition of electron acceleration in this scenario is
\begin{eqnarray}
  \frac{\varv_{\rm dr}}{\varv_\A} \sim \mathcal{M}_\A < \frac{\sqrt{m_r}}{2},
  \label{eq:MA}
\end{eqnarray}
where $\mathcal{M}_\A$ is the {\alf}ic Mach number.
In shocks with a larger {\alf}ic Mach number, magnetic field amplification in the vicinity of the shock front decreases its local {\alf}ic Mach number and once the condition in Equation~\ref{eq:MA} is satisfied, scattering and acceleration of electrons by intermediate-scale unstable wave modes will inevitably proceed.

A similar condition was found in multidimensional kinetic simulations of quasi-perpendicular non-relativistic shocks by \citet{Riquelme+2011}.
Note that this also urges the usage of a realistic mass ratio in non-relativistic shock simulations aimed at studying the dependence of electron acceleration efficiency on the {\alf}ic Mach number.

\subsection{Escape of CRs from acceleration sites}
\label{sec:escape}

After having been accelerated at a shock, the bulk of CRs is advected downstream. In the case of a supernova remnant, this implies CR confinement within the Sedov-Taylor shock radius for most but the highest energy CRs. Towards the end of the Sedov-Taylor phase and the beginning of the snowplow phase, when the shock weakens and/or gets destroyed by the Rayleigh-Taylor instability, CRs can escape the downstream. Because of the strong anisotropy during their escape, CRs likely first excite non-resonant Bell waves that increase the magnetic field and hence the {\alf} speed. As a result, the sub-dominant CR gyro-resonant instability becomes the dominantly growing mode if the CR drift speed drops below $\varv_{\rm dr} < \varv_\A  (16/3 \gamma_i)^{1/3} \alpha^{-2/3}$ (see Section~\ref{sec:bell}). This causes exponential growth of resonant {\alf} waves that scatter CRs and reduce their drift speed further. Hence, in the traditional CCRH picture, the interplay of growing gyro-resonant {\alf} waves and damping processes would modulate CR transport and set the effective escape speed~\citep{Drury2011}.

However, as we have shown here, once the CR drift speed is reduced below $\varv_{\rm dr} < \varv_\A  \sqrt{m_r}/2$, the intermediate-scale instability starts to grow with a rate that is $\sim20$ times that of the gyro-resonant instability (see Fig.~\ref{fig:Intermediate_condition}). This will not only slow down CR propagation and imply steeper CR gradients but also substantially increase their scattering rate so that the effective diffusion coefficient drops and confines CRs more in comparison to the picture in which only the gyro-resonant instability is at work. In summary, we expect an ordering of the dominant instability along the path of escape that looks as follows with increasing distance from the shock: close to the shock, Bell modes are excited (provided the anisotropy is large enough for them to be excited),
followed by the CR gyro-resonant instability, and finally, the intermediate-scale instability after the resonant {\alf} modes have reduced the CR drift speed to values so that it can be excited.

As a result, CRs should be confined longer to the immediate vicinity of their acceleration sites and hadronically interact with the ambient dense interstellar medium so that this new picture would predict brighter gamma-ray halos at GeV and TeV energies in comparison to the traditional picture. These gamma-ray halos surrounding supernova remnants may be observable with the next-generation imaging air-Cherenkov telescopes such as the Cherenkov Telescope Array (CTA).

\section{Conclusion} \label{sec:conclusion}

In this paper we consider linear instabilities that are driven by CRs with a gyrotropically cold momentum distribution.
Solutions of the linear dispersion relation show that CRs with a finite pitch angle drive a dominant intermediate-scale instability between the ion and electron gyroradii, provided the CRs are drifting with a velocity less than half of the {\alf} speed of electrons.
We find that the intermediate-scale instability is typically more than an order of magnitude faster than that at the gyroscale of CR ions.

We provide a classification of various unstable wave modes driven by CR ions at various values of parameters and summarize the conditions for dominance of unstable wave modes (in the linear regime) at various CR drift velocities and pitch angles (see Figure~\ref{fig:regimes} and Table~\ref{table:instabilities}). 

Using an ab inito kinetic simulation, we demonstrate the growth of the dominant intermediate-scale instability and study the non-linear saturation in the simulation.
We show with analytic arguments, which are confirmed by our simulation, that during the initial growth phase the electromagnetic modes are comoving with CRs and exert only work in the parallel direction so that the CR distribution is primarily spread in the parallel direction. As the energy cascades from the most unstable wave length to larger scales, the instability saturates and eventually allows for the excitation of the slower gyroscale instability \citep{Kulsrud_Pearce-1969}.
This is accompanied by an energy-conserving spread of CR velocities in the {\alf} frame of the background plasma.

The intermediate-scale instability is artificially suppressed in the solution of the dispersion relation of gyrotropically cold CRs when the popular assumption $\omega \ll \Omega_i$ is adopted.
This suggests that by considering the contribution of CRs with a power-law momentum distribution without such an assumption, a similar instability may arise.
Two recent kinetic simulations of CRs with power-law momentum distributions that satisfy the criteria of exciting the intermediate-scale instability are presented by \citet{HS+2019}. Both show an initially faster growth similar to what we find in our simulation.

We identify three major areas of research in which the intermediate-scale instability might play a decisive role: 
1.\ CR transport and feedback in galaxies and galaxy clusters, 2.\ electron injection into diffusive shock acceleration, and 3.\ CR escape from the sites of particle acceleration.
\begin{enumerate}
\item The intermediate-scale instability can alter the transport of CRs in galaxies and clusters since the condition of exciting them is typically satisfied. In tandem with the gyro-scale instability, it increases the spectral support and growth rate of electromagnetic background waves which should couple CR tighter to the thermal background plasma. Thus, this facilitates CR feedback by driving powerful winds from galaxies and efficiently heating the cooling cores in galaxy clusters.

\item We show that the intermediate-scale instability can also play a key role for the injection of CR electrons into the process of diffusive shock acceleration at non-relativistic shocks with local {\alf}ic Mach numbers $\mathcal{M}_\A < \sqrt{m_r}/2$ because those are able to efficiently scatter and pre-accelerate thermal electrons.

\item The intermediate-scale instability can also play an important role in the escape of CRs from their source into the interstellar medium by increasing the CR scattering rate and CR confinement to the source vicinity. This should produce gamma-ray halos surrounding CR sources such as supernova remnants that potentially are observable through hadronically produced gamma-ray emission at GeV and TeV energies via the CTA telescope.
\end{enumerate}

\section*{acknowledgement}

M.S., T.T., and C.P. acknowledge support by the European Research Council under ERC-CoG grant CRAGSMAN-646955.

\begin{appendix}

 \section{Density contrast and Alfv\'en speed in various astrophysical plasmas}
\label{app::alphava}
 
In astrophysical plasmas, densities and magnetic field strengths can vary substantially implying a wide range of $\alpha$ and $\varv_\A$. Here, we will provide an overview of these parameters in different astrophysical environments that are of interest for CR driven instabilities.

The interstellar medium is characterized by several co-existing phases in temperature and density.
The cold neutral and molecular phases have temperatures of tens of Kelvin, are mostly composed of neutral atoms and molecules, and have a very low ionization level. Thus, ion-neutral damping strongly depletes {\alf} waves so that CRs are virtually not coupled to the background plasma and thus, this phase is not of interest here.
The warm ionized medium has a temperature of about $10^4$~K and an ion number density of  $\sim 1~{\rm cm}^{-3}$. Assuming equipartition of CR and thermal pressures results in $\alpha\sim10^{-9}$ for GeV CRs \citep{Boulares1990}, and assuming the an average magnetic field strength of $5-15~\mu$G \citep{Beck2015} results in {\alf} speeds of $\varv_\A \sim (3\times 10^{-5} - 10^{-4}) ~c$.
The hot ionized or coronal phase of the interstellar medium is roughly in pressure equilibrium with the warm ionized medium, has a temperature of about $10^6$~K, and an ion number density of $\sim 10^{-2}~{\rm cm}^{-3}$. Assuming a constant CR number density across the different phases yields $\alpha \sim 10^{-7}$.
Adopting magnetic field strengths of $2-4 \mu$G results in {\alf} speeds of $\varv_\A \sim (2 \times 10^{-4} - 4 \times 10^{-4})~c$.
The circum-galactic medium shares many similarities to the hot ionized phase of the interstellar medium regarding these parameters \citep{Tumlinson2017}.

The  intracluster medium in galaxy clusters is in approximate hydrostatic equilibrium with the dark matter-dominated potential. This plasma reaches virial temperatures of $10^7-10^8 $~K and has a typical ion density of $\sim 10^{-3}~{\rm cm}^{-3}$. Assuming that only 1\% of the pressure is provided by GeV CRs (which is inline with upper limits on the gamma-ray emission by \textit{Fermi}, \citealt{Ackermann2014}) results in $\alpha \sim 10^{-8}$.
Faraday rotation measurements suggest root-mean square magnetic field strengths of $2-4~\mu$G \citep{Carilli2002}, which results in {\alf} speeds of $\varv_\A \sim (4 \times 10^{-4} -  10^{-3})~c$.

For supernova remnants, the ion number density of the ambient interstellar medium is $\sim 10^{-1}~{\rm cm}^{-3}$. Assuming a CR energy acceleration efficiency of 10\% and that the shock thermalizes the surrounding plasma to X-ray emitting temperatures of $\sim 10 $~keV, the density ratio of GeV CRs is $\alpha \sim 10^{-6}$. The magnetic field strength varies from $15-150 ~\mu$G~\citep{Reynolds+1981,Aharonian+2005,Parizot+2006}, which results in {\alf} speeds of $\varv_\A \sim (3 \times 10^{-4} - 3\times 10^{-3})~c$.

\begin{figure}
\includegraphics[width=8.7cm]{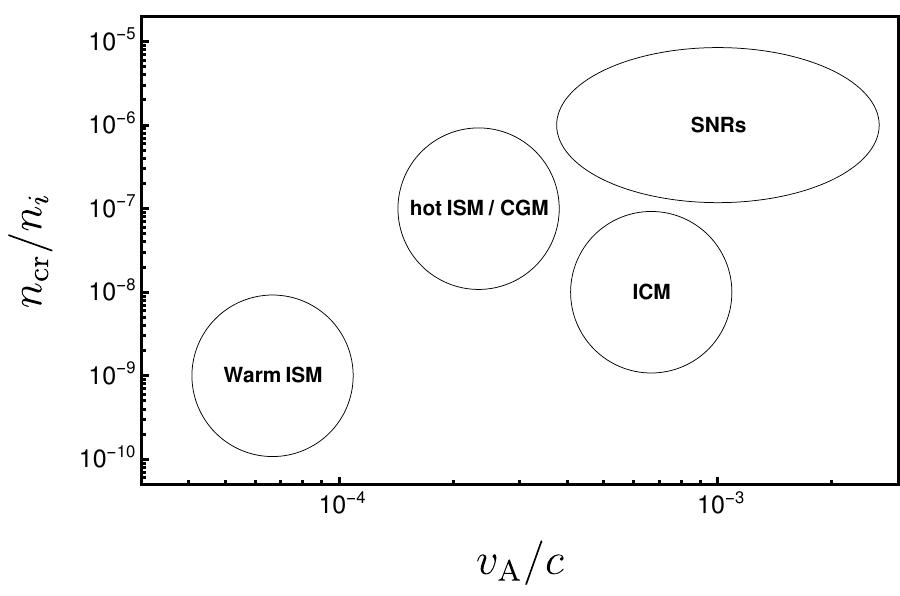}
\caption{
\label{fig:alphava}
Typical values of CR to background ions density ratio ($n_{\rm cr}/n_i$) and {\alf} speed in various astrophysical environments, including the interstellar medium (ISM), the circum-galactic medium (CGM), the intracluster medium (ICM) in galaxy clusters, and supernova remnants (SNRs).
}
\end{figure}

We show an overview of these variables in Figure~\ref{fig:alphava}.
This motivates the choice of $\alpha=10^{-5} \ll 1$ and $\varv_\A/c=10^{-4} \ll 1$ used in our investigation of the solutions of the dispersion relation (Equation~\ref{eq:dispfull}) in Section~\ref{sec:sols}.

\section{Kinetic Linear dispersion relation for magnetized plasmas: parallel modes }
\label{app:dispersion}

The derivation of the linear dispersion relation starts with the Vlasov-Maxwell system of equations that govern the evolution of collision-less magnetized plasmas in phase space.
Assuming an equilibrium configuration with a uniform background magnetic field, $\vec{B}_0$, and no net electric field,\footnote{For this to be an equilibrium configuration, the total current has to be necessarily zero.} we consider the evolution of small perturbations and linearize the system.
The linear dispersion relation of such a system determines the stability of perturbations. In the Fourier space, it determines whether a perturbation with a given wavenumber $\vec{k}$ is stable or not, i.e., whether it exponentially decays or grows in time\footnote{Here, we take the approach of assuming that the wave-numbers $k$ are real and search for complex roots of $\omega$. Thus, evanescent wave modes, i.e., initially traveling disturbances (wave packets) that decay or get amplified as they travel in space, are not included in our discussion here \citep{boyd}.}.

The resulting system that determines the evolution of such a linear perturbation is then solved on unperturbed orbits around the background magnetic field \citep{Krall+1973,Schlickeiser+2002}, and thus the unperturbed phase-space density $f_{s,0}(\vec{u})$ is assumed to be gyrotropic, i.e., $f_{s,0}(\vec{u}) = f_s(u_{\parallel},u_{\perp})$, where the parallel and perpendicular components of the spatial part of the 4-velocity ($u_{\parallel}$ and $u_{\perp}$, respectively) are defined with respect to the direction of the uniform background magnetic field $\vec{B}_0$. The resulting equations can be written as
\begin{eqnarray}
\label{eq:vphEq}
T^{ij} E^j_1 = 0
{\rm, } ~~~~~~~~ 
\vec{k} \times \vec{E}_1
=
\omega \vec{B}_1,
~~~~ {\rm and } ~~~~
\vec{k} \cdot \vec{B}_1 = 0,
\end{eqnarray}
where $T^{ij}$ is the dispersion tensor, $\vec{E}_1$ and $\vec{B}_1$ are the linear perturbations of the electric and magnetic fields in the Fourier-space, respectively, labeled by wavenumber $\vec{k}$. These perturbations oscillate in time with a frequency $\omega$. We obtain the last  two equations by linearizing Maxwell's equations.

The dispersion tensor, $T^{ij}$, is very complicated, see, e.g., Equations (8.3.53)-(8.3.57) of \citet{Schlickeiser+2002}. However, it is substantially simplified if we  only consider wave modes that are parallel to $\vec{B}_0$\footnote{For CR-induced instabilities, modes that are not parallel to $\vec{B}_0$ are typically subdominant in their growth with respect to parallel modes \citep{Zweibel2017,Kulsrud_Pearce-1969}.}.
For these parallel modes, i.e., $\vec{B}_0  = B_0 \bs{\hat{\vec{x}}} $ and $\vec{k}   = k ~ \bs{\hat{\vec{x}}} $, the dispersion tensor has the simple form
\begin{eqnarray}
T^{ij}
=
\left(
\begin{array}{ccc}
T^{11} & 0 & 0 \\
 0 &T^{22} &T^{23} \\
 0 & -T^{23} &T^{22} \\
\end{array}
\right).
\end{eqnarray}
The explicit form of the matrix elements are given below and thus the dispersion relation is the determinant of the dispersion tensor, $|T^{ij}| = T^{11} \left[  (T^{22})^2 + (T^{23})^2 \right]=0$.
That is, we find two different type of solutions which we detail below.

\subsection{Electrostatic modes: $T^{11}=0$ }

In this case, only the parallel electric field is finite, i.e., $E_{x,1} \neq 0$, while the perpendicular components of the electric and magnetic field are zero, i.e., $E_{y,1} = E_{z,1} = 0  \Rightarrow B_{y,1} = B_{z,1} = 0$. In this case the dispersion relation is given by
\begin{eqnarray}
T^{11}
&=&
1
-
 \sum_s \frac{ q^2_s }{m_s  } 
\int d^3u
 \frac{f_{s,0}  }{\gamma}  
\frac{  (1-\varv_x^2 /c^{2})}
{\left(\omega -k  \varv_x\right){}^2}
=0.
~~~~~
\end{eqnarray}

\subsection{Electromagnetic modes: $T^{22} \pm i T^{23} =0$ }

For this solution, $E_{x,1} = 0$, but $E_{y,1} \neq 0$ and $ E_{z,1} \neq 0$, which implies $B_{y,1} \neq 0$ and $B_{z,1} \neq 0$.
The dispersion relation of these electromagnetic modes, $D^{^{\pm}} $, can be written as
\begin{eqnarray}
D^{^{\pm}} = T^{22} \pm i T^{23} =    1 -\frac{k ^2c^2}{\omega^2} + \sum_s \chi^{\pm}_s = 0,
\end{eqnarray}
where $\chi_s^{\pm}$ is the linear response for species, $s$, with charge $q_s$, mass $m_s$ and an equilibrium gyrotropic phase-space distribution $f_{s,0}$.
We can write  $\chi^{\pm}_s$ in several forms as follows.

\begin{align}
\chi_s^{\pm}
&=
\frac{ 1 }{\omega^2} 
\frac{ q^2_s }{m_s  } 
\!\int\!\! d^3u
 \frac{f_{s,0}  }{\gamma}
\left[
\frac{\omega -k \varv_x}{k \varv_x-\omega \pm \Omega _s }
-
\frac{\varv_{\perp}^2 c^{-2}\left(k^2c^2-\omega ^2  \right)}
{2 \left(k \varv_x - \omega  \pm \Omega_s\right){}^2}
\right]
\nonumber 
\\
&=
\frac{ 1 }{\omega^2} 
\frac{ q^2_s }{m_s  } 
\!\int\!\! d^3u
\frac{f_{s,0}  }{\gamma}
\left[
\frac{\omega -k \varv \mu}{k \varv  \mu -\omega \pm \Omega _s }
-
\frac{\varv^2 (1- \mu^2) \left(k ^2 c^2-\omega ^2  \right)}
{2 c^2 \left(k  \varv  \mu - \omega  \pm \Omega_s\right){}^2}
\right]
\nonumber 
\\
&= 
\frac{ -1 }{ 2 \omega }  
\frac{q_s^2}{m_s } 
\!\int\!\! d^3u
\frac{ \varv(1-\mu^2)}{k  \varv \mu -\omega \pm \Omega _s } 
\left[
\partial_{u} f_{s,0}
+
\left(  \frac{ k  \varv }{ \omega } - \mu \right)
\frac{ \partial_{ \mu} f_{s,0} }{u } 
\right]
\nonumber \\
&=
\frac{ - 1}{ 2 \omega^2 }  
\frac{q_s^2}{m_s } 
\!\int\!\! d^3u
\frac{ u_{\perp}^2}{\gamma^2} 
\left[
\frac{ 
\left[ \left( \omega -  k  \varv_x  \right) /\varv_{\perp} \right] \partial_{ u_{\perp}} f_{s,0} 
+
k_{\parallel}  \partial_{u_x} f_{s,0}
}{k  \varv_x -\omega  \pm \Omega _s }
\right],
\label{eq::disp04}
\end{align}
where $\vec{u} =\gamma \bvarv$ is the spatial part of the 4-velocity, $u_{\parallel}$ and $u_{\perp}$  are the parallel and perpendicular velocity, respectively, which are defined with respect to the direction of the uniform background magnetic field $\vec{B}_0$.
The magnitude of the velocity is defined such that
$u^2 = u_{\parallel}^2 + u_{\perp}^2  $, and $\mu = u_{\parallel} /u $ is the pitch angle.
The momentum-space integrals can be replaced by
$$
\int d^3u = 2 \pi \int_0^{\infty} u^2 du \int_{-1}^{1} d\mu = 2 \pi 
\int_0^{\infty}  u_{\perp} du_{\perp} \int_{-\infty}^{\infty} du_x~~.
$$
The non-relativistic cyclotron frequency of the species $s$ is
$\Omega_{s,0}  = q_s B_0/ m_s   $, and $\Omega_s =  \Omega_{s,0} / \gamma $ is the relativistic cyclotron frequency of a particle with Lorentz factor $\gamma$.
The solutions of $D^{\pm}=0$ correspond to different polarization of the perturbed electric and magnetic fields. 
That is, $D^{\pm} =0$ corresponds to modes with right/left polarization such that $ E_{y,1}(k,\omega) \pm i E_{z,1}(k,\omega) =0$, and $ B_{y,1}(k,\omega) \pm i B_{z,1}(k,\omega) =0$.

\subsubsection{Symmetry of the dispersion relations of electromagnetic wave modes}
\label{app:symmetry}

The electromagnetic dispersion relation $D^{\pm}$ exhibits a remarkable symmetry that tells us important aspects about the nature of their solutions.
All quantities in the dispersion relations are real numbers except for $\omega$, which is a complex function. Thus, if $ \omega=\omega_s$ is a solution of $D^{+}(k=k_s)=0$, then the following holds true.
\begin{enumerate}

\item The complex conjugate of $\omega_s$, i.e., $ \omega=\omega_s^{\star}$ is also a solution of $D^{+}(k=k_s)=0$.
The same is also true for the solutions of $D^{-}=0$.

\item At $k=-k_s$, the dispersion relation $D^{-}(k=-k_s)=0$ has the two solutions $\omega= - \omega_s$ and $ \omega=  - \omega_s^{\star}$.

\end{enumerate} 

These two properties imply that if a polarization state at $k=k_s$ has a fastest-growing wave mode that grows with the rate $\Gamma$ and oscillates with the rate $\omega_r = \mathrm{Re}[\omega_s]$, then the opposite polarization state at $k= -k_s$ has its fastest-growing wave mode that grows with the same rate $\Gamma$, but oscillates with the rate $\omega_r = -\mathrm{Re}[\omega_s]$.  
Thus, both unstable modes with different polarization states have the same phase velocity. Hence, the growing perturbed electric and magnetic fields of both modes rotate in the same direction around $\vec{B}_0$.

\subsubsection{Scattering in velocity space by an electromagnetic wave mode}
\label{app:scattering}

To compute the impact of an electromagnetic wave on particle velocities, we consider the non-relativistic Lorentz force on a particle of species $s$ with change $q_s$ and mass $m_s$ for simplicity. In Fourier space, we obtain\footnote{Formally,  the right hand side of equation~\eqref{eq:lfk} should be integrated over $k$, and we drop this because we only consider the scattering by a single mode.}
\begin{eqnarray}
\label{eq:lfk}
\frac{d \bvarv}{dt} = \frac{q_s}{m_s} \left( \vec{E}_k + \bvarv \times \vec{B}_k \right).
\end{eqnarray}
Below we omit the subscript $k$ to avoid unnecessary cluttering of subscripts.
We use this expression to compute the rate of change of kinetic particle energies parallel and perpendicular to the background magnetic field direction.
The rate of change of the parallel energy of a particle is
\begin{eqnarray}
\K_{\parallel}
=
\frac{m_s}{2} \frac{d \varv^2_x }{dt}
=
q_s ( \varv_x   E_x + \varv_x \varv_y  B_z - \varv_x  \varv_z B_y ).
\end{eqnarray}
The rate of change of the perpendicular energy of a particle is 
\begin{eqnarray}
\K_{\perp}
&=&
\frac{m_s}{2} \frac{d \varv^2_{\perp} }{dt}
=
\frac{m_s}{2} \frac{d (\varv_y^2 + \varv_z^2) }{dt}
\nonumber \\
&=&
q_s ( \varv_y  E_y + \varv_z  E_z - \varv_x \varv_y  B_z + \varv_x  \varv_z B_y ),
\end{eqnarray}
where we used $\bvarv \cdot \left( \bvarv \times \vec{B}_k \right)=0$.

In Fourier space, the linearized electromagnetic wave modes obey the relations $E_x=0$, $E_y = \varv_{\rm ph} B_z$, and $E_z =-   \varv_{\rm ph} B_z$.\footnote{This is also true for non-linear {\alf} modes because these are exact solutions of Maxwell's equations in the limit $\varv_\A \ll c$.} Thus, the parallel and perpendicular rates of change of particle energy are given by
\begin{eqnarray}
\K_{\parallel}
&=&
\frac{m_s}{2} \frac{d \varv^2_x }{dt}
=
q_s \varv_x ( \varv_y  B_z - \varv_z B_y )
\\
\K_{\perp}
&=&
q_s ( \varv_y  \varv_{\rm ph} B_z - \varv_z \varv_{\rm ph} B_z - \varv_x \varv_y  B_z + \varv_x  \varv_z B_y ),
\nonumber \\
&=&
- q_s \left[ (\varv_x - \varv_{\rm ph} )\varv_y  B_z   -( \varv_x - \varv_{\rm ph} )  \varv_z B_y \right].
\end{eqnarray}

That is, the relative scattering of the particles' velocities by an excited mode (labeled by $k$) depends on its phase speed, $\varv_{\rm ph}$.

\section{SHARP-1D3V}
\label{app:1D3V}

SHARP-1D3V is an extended version of SHARP-1D, i.e., a solver in one spatial and one velocity-space dimension (1D1V) presented by~\cite{sharp}.
As we see below, the reduced equations in the 1D3V case (i.e., one spatial and three velocity-space dimensions) are reminiscent to that in the 1D1V case, with the addition of evolution equations for the perpendicular field and velocity components.
Below we derive such a 1D3V model from the Vlasov-Maxwell system to highlight the various assumptions that enter into such a model. We then show how SHARP-1D is extended to model the 1D3V evolution equations to constitute SHARP-1D3V. 
We have successfully performed various validation tests, including linear Landau damping of Langmuir waves, electromagnetic wave propagation, and \alf-wave propagation using SHARP-1D3V. 
However we postpone a presentation of these tests to a future publication.

\subsection{1D3V model}

The evolution of collisionless plasmas in phase space is governed by the Vlasov-Maxwell equation.
To get the 1D3V electromagnetic version of such a system of equations, we only allow for variations in the $x$ direction and average out the dynamics  in the $y$ and $z$ directions.
Formally we proceed with the following steps to derive such a model.

1. We define the 1D3V version of various quantities in terms of their 3D3V version as
\begin{eqnarray}
F^{\text{1D3V}}(x,t)
&\equiv& \iint \frac{ dy dz}{A_{yz}} ~ F^{\text{3D}}(x,y,z,t),
\end{eqnarray}
where $A_{yz}$ is the total area of the $(y,z)$ plane, i.e., $\iint dy dz = A_{yz}$, and $F$ denotes any component of $\vec{B},~\vec{E}, \text{and~} \vec{J}$ or $\rho$.

For example, the $z$ component of the current density in 3D  is given by
\begin{eqnarray}
J_z(x,y,z,t) 
&=& \sum_{\alpha} q_{\alpha}~ \varv_{z,\alpha} S(x-x_\alpha)~ S(y-y_\alpha)~S(z-z_\alpha), ~~~
\end{eqnarray}
where the sum extends over all plasma macro-particles $\alpha$ at position $\{x_\alpha(t),y_\alpha(t),z_\alpha(t)\}$ and $S$ is the shape of the macro-particle, see, e.g., Appendix B of \cite{sharp}. In 1D3V, this reduces to
\begin{eqnarray}
J^{\text{1D3V}}_z
&=& J_z(x,t) =
\iint \frac{ dy dz}{A_{yz}} J_z(x,y,z,t)
=
\sum_{\alpha} \frac{q_{\alpha}}{A_{yz}} \varv_{z,\alpha} ~ S(x-x_\alpha(t))
\nonumber \\ 
&=& 
\sum_{\alpha} Q^{\text{1D3V}}_{\alpha}  \varv_{z,\alpha} ~ S(x-x_\alpha(t))
 ,
\end{eqnarray}
where $Q^{\text{1D3V}}_{\alpha} = q_{\alpha}/A_{yz} $ is the surface charge density which is taken to be the charge of the particle $\alpha$ in our 1D3V model.

2. We assume that the phase space distribution function of all species is independent of the $y$ and $z$ coordinates in phase space, i.e.,
\begin{eqnarray}
f(x,y,z,\vec{u},t) \equiv 
\frac{ f^{\text{1D3V}}(x,\vec{u},t) }{ A_{yz} }.
\end{eqnarray}
Therefore, the Vlasov equation can be written as
\begin{eqnarray}
0 &=& \partial_t f + \bvarv \cdot  \nabla f
+ \vec{a} \cdot \nabla_{\vec{u}} f  
\nonumber \\
&=& 
\frac{  \partial_t f^{\text{1D3V}} }{A_{yz}}
+ 
\frac{ \varv^x \partial_x  f^{\text{1D3V}} }{A_{yz}}
+ 
\frac{ \vec{a}(x,y,z,t) }{A_{yz}} \cdot \nabla_{\vec{u}} f^{\text{1D3V}} ,
\label{eq:phasespace1d3v}
\end{eqnarray}
where $\vec{a}(x,y,z,t) = (q_s/m_s) (\vec{E} + \bvarv \times \vec{B}) $ is the acceleration due to the Lorentz force on a particle at position $(x,y,z,u_x,u_y,u_z)$ in phase space.
By integrating equation~\eqref{eq:phasespace1d3v} over $y$ and $z$,
we get
\begin{eqnarray}
0 &=& 
\partial_t f^{\text{1D3V}} 
+ 
\varv^x \partial_x  f^{\text{1D3V}}
+ 
\vec{a}^{\text{1D3V}}(x,t)  \cdot \nabla_{\vec{u}} f^{\text{1D3V}}.
\label{eq:phasespace}
\end{eqnarray}
Therefore, in 1D3V model, Equation~\eqref{eq:phasespace} governs the plasma evolution in phase space, i.e., it is the equivalent of the Vlasov equation in 3D3V.

Maxwell's equations in 1D3V imply that $B_x$ has to be constant and uniform, i.e., $\partial_t B_x =0 $ and  $\partial_x B_x=0$. The remaining evolution equations are given by 
\begin{align}
\label{eq:MEs-1D3V}
& 
\partial_x E_x  = \frac{\rho}{\epsilon_0}	,	 	&
& 
0 = \frac{J_x}{\epsilon_0}  +  \partial_t E_x ,		&
\nonumber \\
&
 \partial_t B_y =    \partial_x  E_z   	,	&
& 
  \partial_t B_z = -  \partial_x   E_y   	,	&
\nonumber \\
&
   \partial_t E_y  =  - c^2 \partial_x B_z -   \frac{J_y}{\epsilon_0} 	,	&
&
     \partial_t E_z 	= c^2 \partial_x B_y   - \frac{J_z}{\epsilon_0} 	.&
\end{align}
The charge density and charge current density are given by
\begin{eqnarray}
\rho(x,t) 
&=& \sum_{\alpha} Q_{\alpha} ~ S(x-x_\alpha(t)),
\nonumber \\
\vec{J}(x,t) 
&=& 
\sum_{\alpha} Q_{\alpha} ~ \bvarv_{\alpha}  S(x-x_\alpha(t)).
\end{eqnarray}

In 1D3V, the equations of motion for a macro-particle $\alpha$ with mass $m_{\alpha}$ and charge $q_{\alpha}$ reduce to
\begin{eqnarray}
\frac{d x_{\alpha} }{dt} 
&=&
\varv_{x,\alpha},
\nonumber \\
\frac{d(\gamma_{\alpha} \varv_{x,\alpha}) }{dt} 
&=&
\frac{q_{\alpha} }{ m_{\alpha}} 
\left( 
E_{x,\alpha} +  \varv_{y,\alpha} B_{z,\alpha} -  \varv_{z,\alpha}   B_{y,\alpha} 
\right),
\nonumber \\
\frac{d(\gamma_{\alpha} \varv_{y,\alpha}) }{dt} 
&=&
\frac{q_{\alpha} }{ m_{\alpha}} \left( E_{y,\alpha} +
  \varv_{z,\alpha}   B_{x,0}
- 
  \varv_{x,\alpha}   B_{z,\alpha} \right),
\nonumber \\
\frac{d(\gamma_{\alpha} \varv_{z,\alpha}) }{dt} 
&=&
\frac{q_{\alpha} }{ m_{\alpha}} \left( E_{z,\alpha} + 
  \varv_{x,\alpha}   B_{y,\alpha} 
-
 \varv_{y,\alpha}  B_{x,0}
 \right),
\label{eq:EOM-1D3V}
\end{eqnarray}
where $\vec{E}_{\alpha}$ and $\vec{B}_{\alpha}$ are the electromagnetic fields at the position of the macro-particles $\alpha$, centered on $x_{\alpha}$. These equations of motion are obtained by taking 4-velocity moments of the reduced Vlasov Equation~\eqref{eq:phasespace} \citep{birdsall+1980}.

\subsection{Discretized evolution equations of SHARP-1D3V}

The normalization of the evolution equations follows the presentation in \cite{sharp}. That is, all quantities below are dimensionless and are normalized to code units.
We have the freedom to define $\rho$, $J_y$, and $J_z$ on our 1D grid. We choose to define them at the grid centers, i.e., at $x_{i+1/2}$ for the $i$th cell, which automatically defines the positions of the other field components to achieve second order accuracy both in space and time (see Table~\ref{tab:1d3v-grid}).
The interpolation from particles to charge and current densities on the grid is done as follows.

\begin{eqnarray}
J ^{n-\frac{1}{2}}_{y,i+\frac{1}{2}}
&=&
\frac{1}{N_{pc}}
\sum_{\alpha} \bar{Q}_{\alpha} ~  \varv^{n-1/2}_{y,\alpha} ~ 
W^m\left(  x ^{n}_{\alpha} - \frac{\Delta  t }{2}  \varv ^{n-1/2}_{x,\alpha} -  x _{i+1/2}\right),
~~~~~~~~~~
\\
J ^{n-\frac{1}{2}}_{z,i+\frac{1}{2}}
&=&
\frac{1}{N_{pc}}
\sum_{\alpha} \bar{Q}_{\alpha} ~  \varv ^{n-1/2}_{z,\alpha} ~ 
W^m\left( x ^{n}_{\alpha} - \frac{\Delta  t }{2}  \varv ^{n-1/2}_{x,\alpha} -  x _{i+1/2}\right),
\\
\rho ^{n}_{i+\frac{1}{2}}
&=&
\frac{1}{N_{pc}}
\sum_{\alpha} \bar{Q}_{\alpha} ~
W^m\left( x ^{n}_{\alpha} -  x _{i+1/2}\right),
\end{eqnarray}
where $N_{pc} = \left( \sum_s \bar{Q}^2_s N_s/ \bar{M}_s\right) /N_c $, and $\bar{Q}_s$ and $\bar{M}_s$ are the  normalized mass and charge, respectively, of a species with $N_s$ particles on a grid with $N_c$ cells.
The weight function, $W$, is a function that depend on the order of the shape of macro-particles (see Appendix B of \citealt{sharp}).

The computational domain is divided into $N_c$ cells with constant size $\Delta x$, and we update the variables forward in time  with the time step $\Delta t$.
Thus, the discretized Maxwell's equations  (in  space and time with $O(\Delta x^2)$ and  $O(\Delta t^2)$ accuracy) can be divided up into equations similar to the 1D1V case  \citep{sharp},
\begin{eqnarray}
E^{n}_{x,i+1}  &=& E^{n}_{x,i} + \Delta x ~ \rho ^n_{i+\frac{1}{2}} ,	
\\
E^{n}_{x,\text{tot}} &\equiv&
\sum_i E^{n}_{x,i+1}
=
E^{n-1}_{x,\text{tot}}
+ \Delta  t \sum_{\alpha} q_{\alpha} \varv^{n-\frac{1}{2}}_{x,\alpha},
\end{eqnarray}

and the transverse evolution equations read
\begin{eqnarray}
E^{n}_{y,i+\frac{1}{2}} 
&=&
E^{n-1}_{y,i+\frac{1}{2}}
-  \Delta  t ~  J ^{n-\frac{1}{2}}_{y,i+\frac{1}{2}}
-
 \frac{ \Delta  t}{\Delta x} 
\left\{
B^{n-\frac{1}{2}}_{z,i+1} - B^{n-\frac{1}{2}}_{z,i}   		
\right\},
\\
E^{n}_{z,i+\frac{1}{2}} 
&=&
E^{n-1}_{z,i+\frac{1}{2}}
-  \Delta  t ~  J ^{n-\frac{1}{2}}_{z,i+\frac{1}{2}}
+
 \frac{ \Delta  t}{\Delta x} 
\left\{
B^{n-\frac{1}{2}}_{y,i+1} - B^{n-\frac{1}{2}}_{y,i}   		
\right\},
\\
B^{n+\frac{1}{2}}_{y,i} 
&=&
B^{n-\frac{1}{2}}_{y,i}
+ \frac{ \Delta  t}{\Delta x} 
\left\{
E^{n}_{z,i+\frac{1}{2}} - E^{n}_{z,i-\frac{1}{2}}   		
\right\},
\\
B^{n+\frac{1}{2}}_{z,i} 
&=&
B^{n-\frac{1}{2}}_{z,i}
- \frac{ \Delta  t}{\Delta x} 
\left\{
E^{n}_{y,i+\frac{1}{2}} - E^{n}_{y,i-\frac{1}{2}}   		
\right\}.
\end{eqnarray}
The grid assignments and  the time staggering of the various quantities are shown in Table~\ref{tab:1d3v-grid}.

In order to evolve the particles, we need to interpolate value of the fields from the grid onto the macro-particles. For a macro-particle at $x_{\alpha}$ that has a weight function $W_{i,x_{\alpha}} \equiv W(x_i-x_{\alpha})$, this is done as follows

\begin{eqnarray}
E^x_{\alpha}
&=&
\sum_i
E^x_i ~ W_{i,x_{\alpha}},
\nonumber \\
E^y_{\alpha}
&=&
\sum_i
E^y_{i+\frac{1}{2}} ~ W_{i+\frac{1}{2},x_{\alpha}}
{,}~~~~~
E^z_{\alpha}
=
\sum_i
E^z_{i+\frac{1}{2}} ~ W_{i+\frac{1}{2},x_{\alpha}},
\nonumber \\
B^y_{\alpha}
&=&
\sum_i
B^y_{i} ~ W_{i ,x_{\alpha}}
{,}~~~~~~~~~~~~
B^z_{\alpha}
=
\sum_i
B^z_{i} ~ W_{i ,x_{\alpha}}.
\end{eqnarray}

To update the macro-particles' momenta we define $\vec{B}^{n}_{\alpha} = 
\left( 
 \vec{B}^{n+\frac{1}{2}}_{\alpha} + \vec{B}^{n-\frac{1}{2}}_{\alpha}
\right)/2$, which implies

\begin{eqnarray}
\label{eq:EOM-1D3V-2}
\vec{u}^{n+\frac{1}{2}}_{\alpha}
= 
\vec{u}^{n-\frac{1}{2}}_{\alpha}
+
\frac{q_{\alpha} \Delta t }{ m_{\alpha}} 
\left[ 
\vec{E}^{n}_{\alpha} + \frac{1}{2}
\left(
\frac{\vec{u}^{n+\frac{1}{2}}_{\alpha}}{\gamma^{n+\frac{1}{2}}_{\alpha}}
+
\frac{\vec{u}^{n-\frac{1}{2}}_{\alpha}}{\gamma^{n-\frac{1}{2}}_{\alpha}}
 \right)
\times \vec{B}^n_{\alpha} 
\right].
&& ~~~~~~~
\end{eqnarray}
That is, the particle's trajectory is evolved by using the relativistic Vay pusher \citep{Vay-2008}.
This can be schematically represented as
\[
\{
\vec{u}^{n+\frac{1}{2}}_{\alpha}
~~ \& ~~
\gamma^{n+\frac{1}{2}}_{\alpha}
~~ \& ~~
\vec{E}^{n}_{\alpha}
~~ \& ~~
\vec{B}^{n}_{\alpha}
\}
\xRightarrow[]{  \text{Vay\_Pusher()} ~ } 
\{
\vec{u}^{n+\frac{1}{2}}_{\alpha}
~~ \& ~~
\gamma^{n+\frac{1}{2}}_{\alpha}
\}
\]

The macro-particles' positions are then updated via
\begin{eqnarray}
x^{n+1}_{\alpha} 
=
x^{n}_{\alpha} 
+ \Delta  t \frac{u ^{n+\frac{1}{2}}_{x,\alpha}
}{
\gamma^{n+\frac{1}{2}}_{\alpha}
}
.
\end{eqnarray}

\begin{deluxetable}{ ccc }[h!]
\tablewidth{8cm}
\tabletypesize{\footnotesize}
\tablecolumns{4} 
\tablecaption{
The grid assignment and time staggering  of the  SHARP-1D3V code.  \label{tab:1d3v-grid} 
}
\tablehead{
&  $x_{i-1}$ & $x_{i-\frac{1}{2}}$  \vspace{0.1cm}  ~  }
\startdata
\rule{0pt}{8pt}
$t^{n-1}$	&  $\{E_x\}$ & $\{\rho , E_y ,E_z\}$ ~
\vspace{0.2cm} 
~
\\
 $t^{n-\frac{1}{2}}$	&   $\{B_y,B_z\}$ 	&  $\{J_y,J_z\}$ 
\vspace{0.2cm} 
~
\enddata
\end{deluxetable}

\end{appendix}

\bibliography{refs}
\bibliographystyle{aasjournal}

\end{document}